\documentclass[12pt]{article}
\pdfoutput=1
\usepackage{sectsty}
\usepackage[toc,page]{appendix}
\usepackage{mathrsfs}
\usepackage{mathtools}
\usepackage{tikz-cd}
\allsectionsfont{\sffamily}
\newcommand{\x}{{\rm x}}

\newcommand{\dd}{{\rm d}}
\newcommand{\expe}{{\rm e}}

\usepackage[nosort]{cite}
\usepackage{enumerate}

\usepackage[font={scriptsize,singlespacing}]{caption}
\usepackage{epstopdf}

\usepackage{amsmath}
\usepackage{amssymb}
\usepackage{subfigure}
\usepackage{hyperref}
\usepackage{url}
\usepackage{xcolor}
\usepackage{color}
\definecolor{amaranth}{rgb}{0.9, 0.17, 0.31}
\definecolor{purple(munsell)}{rgb}{0.62, 0.0, 0.77}
\definecolor{americanrose}{rgb}{1.0, 0.01, 0.24}
\definecolor{palatinateblue}{rgb}{0.15, 0.23, 0.89}
\definecolor{royalblue(web)}{rgb}{0.25, 0.41, 0.88}
\definecolor{hanpurple}{rgb}{0.32, 0.09, 0.98}
\definecolor{beaublue}{rgb}{0.74, 0.83, 0.9}
\definecolor{carminered}{rgb}{1.0, 0.0, 0.22}
\definecolor{brightpink}{rgb}{1.0, 0.0, 0.5}
\definecolor{vividviolet}{rgb}{0.62, 0.0, 1.0}
\hypersetup{ linktoc=all,
    colorlinks, linkcolor={palatinateblue},
    citecolor={brightpink}, urlcolor={amaranth}}

\def\sideremark#1{\ifvmode\leavevmode\fi\vadjust{\vbox to0pt{\vss% the remark
 \hbox to 0pt{\hskip\hsize\hskip1em%                          will appear only
 \vbox{\hsize2cm\tiny\raggedright\pretolerance10000%          on the side
 \noindent #1\hfill}\hss}\vbox to8pt{\vfil}\vss}}}%
\newcommand{\bo}{\raise-1mm\hbox{\Large$\Box$}}

\newcommand{\be}{\begin{equation}}
\newcommand{\ee}{\end{equation}}
\newcommand{\bea}{\begin{eqnarray}}
\newcommand{\eea}{\end{eqnarray}}

\begin{document}
\thispagestyle{empty}
\begin{center}

\null \vskip-1truecm \vskip2truecm
%{\LARGE{\bf \textsf{The effective temperatures of stationary Unruh-DeWitt detectors}}}
{\LARGE{\bf \textsf{Unruh-like effects: Effective temperatures along stationary worldlines}}}

\vskip1truecm
\textbf{\textsf{ Michael Good${}^{1}$, Benito A. Ju\'arez-Aubry${}^{2}$, Dimitris  Moustos${}^{3}$, Maksat Temirkhan${}^{1}$}}\\
%Nosratollah Jafari
%{\footnotesize\textsf{Nazarbayev University, Astana, Kazakhstan}\\
{\footnotesize\textsf{
${}^1$Department of Physics, Nazarbayev University,\\
Nur-Sultan 010000, Kazakhstan\\
${}^2$Departmento de F\'isica Matem\'atica, Instituto de Investigaciones en Matem\'aticas Aplicadas y en Sistemas, Universidad Nacional Aut\'onoma de M\'exico,\\
Mexico City 20126,  Mexico\\
${}^3$Department  of  Physics,  University  of  Patras, \\
Patras 26504,  Greece}\\
%{\tt Email: michael.good@nu.edu.kz}}\\
{\tt Email:  michael.good@nu.edu.kz, benito.juarez@iimas.unam.mx, dmoustos@upatras.gr, maksat.temirkhan@nu.edu.kz }}\\

\end{center}
\vspace{0.4cm}
%\vskip1truecm 
\centerline{\textsf{ABSTRACT}} \baselineskip=15pt

\medskip

\vskip0.4truecm
%\hrule
%\end{abstract}
%\pacs{03.70.+k, 04.62.+v, 04.60.-m} %04.62.+v is quantum fields in CS, and 03.70.+k is quantized fields, 04.60.-m is quantum gravity
%\keywords{dynamical Casimir effect, moving mirrors, negative energy flux}%Use showkeys class option if keyword
%\maketitle
%\tableofcontents
%\newpage
\vspace{0.4cm}

%\dimitris{I still believe the abstract is too long. Trying to shorten it a bit:} %%
We study the detailed balance temperatures recorded along all classes of stationary, uniformly accelerated worldlines in four-dimensional Minkowski spacetime, namely along (i) linear uniform acceleration, (ii) cusped, (iii) circular, (iv) catenary, and (v) helix worldlines, among which the Unruh temperature is the particular case for linear uniform acceleration.  As a measuring device, we employ an Unruh-DeWitt detector, modeled as a qubit that interacts for a long time with a massless Klein-Gordon field in the vacuum state. The temperatures in each case (i) - (v) are functions of up to three invariant quantities: curvature or proper acceleration, $\kappa$, torsion, $b$, and hypertorsion, $\nu$, and except for the case (i), they depend on the transition frequency difference of the detector, $\omega$. We investigate numerically the behavior of the frequency-dependent temperatures for different values of $\kappa$, $b$, and $\nu$ along the  stationary worldlines (ii) - (v) and evaluate analytically the regimes where the temperatures recorded along the different worldlines coincide with each other in terms of relevant asymptotic limits for $\kappa$, $b$, or $\nu$, and discuss their physical meaning.  We demonstrate that the temperatures in cases (ii) - (v) dip under the Unruh temperature at low frequencies and go above the Unruh temperature for large $|\omega|$. It is our hope that this study will be relevant to the design of experiments seeking to verify the Unruh effect or generalizations thereof.

\vspace{1cm}
\noindent
\emph{With gratitude to all essential workers during the} COVID-19 \emph{pandemic.}

\section{Introduction}

%\benito{The following two paragraphs will be moved to the introduction:}

%The instant simple-minded connection to Hawking radiation \cite{hawking} is heuristically made by elementary substitution of the Newtonian surface gravity acceleration $g = GM/R^2$ at the Schwarzschild radius $R= 2GM/c^2$,
%\be T_H = \frac{\hbar c^3}{8\pi G M k_B}. \ee

%\benito{Paragraphs end here.}

Linearly uniformly accelerated observers in Minkowski spacetime perceive the Minkowski vacuum state as a thermal state at a temperature proportional to their acceleration, known as the Unruh temperature. The Unruh effect \cite{Unruh:1976db, Unruh:review, Earman} constitutes one of the most remarkable theoretical predictions of quantum field theory, which comes about from investigating the lessons learned in curved spacetimes applied to Minkowski spacetime.

The experimental verification of the Unruh effect has however remained elusive so far. The physical Unruh temperature carries a small proportionality factor that makes experiments technically challenging. If $\hbar$, $c$ and $k_{\rm B}$ are the reduced Planck constant, the speed of light and Boltzmann's constant respectively, the Unruh temperature is
\be T_{\rm U} = \frac{\hbar \kappa}{2\pi c k_{\rm B}},
\label{UnruhPhys}
\ee
where $\kappa$ is the proper acceleration of the observer, and one realizes that for values of $\kappa$ of order $\sim 10^{20}\textrm{m}/\textrm{s}^2$ the Unruh temperature is of the order of $\sim 1$K. This makes it clear that detectors or laboratory instruments must highly accelerate for any effect to be measured above noise level, even in ultracold environments.

The investigation of circular acceleration has been posed as a viable experimental alternative instigated by Bell and Leinaas \cite{Leinaas.Bell,Bell:Leinaas} and later by Unruh \cite{Unruh:circ}. In this case, the temperature measured in the experiment would not be constant, but should depend (in addition to the details of the curved motion) on the typical transition frequency difference of the probe (e.g., an atom) or detector used as a measurement device. Thus, this is a generalized Unruh temperature, $T_{\rm cir}(\omega)$, where $\omega$ is the transition frequency difference.

The sense in which such frequency-dependent function deserves to be called a temperature is discussed in detail in \cite{BD} by two of us (see also Sec. \ref{thermalize} below), making use of a generalization of the detailed balance condition for Unruh-DeWitt detectors (see \cite{FBL}) with a frequency-dependent function -- the frequency-dependent temperature -- instead of a constant temperature. The point is that detectors that interact for a long time with a quantum field will reach a Gibbs state at late times at such frequency-dependent, effective \emph{temperature}, if such generalized detailed balance condition holds.  With respect to such a circular Unruh-like effect, a concrete experimental proposal for its measurement was proposed in \cite{BillTalk} and a paper containing the development of such ideas will appear in the near future. The theoretical basis for such experimental proposal will be appear in \cite{Gooding:2020}.

On the other hand, constant-acceleration, stationary motions in Minkowski spacetime have been classified since the work of Letaw \cite{Letaw:1980yv}. In addition to the well-known linear uniform and circular acceleration motions, three additional non-trivial (i.e., with acceleration $\kappa \neq 0$) uniform acceleration motions, generated by linear combinations of Killing vector fields (KVF), define stationary trajectories in Minkowski spacetime. Such trajectories are classified up to isometries in terms of three constants of motion of the curve traced by the trajectory in spacetime, namely proper acceleration or curvature, $\kappa$, torsion, $b$, and hypertorsion, $\nu$, and are named either in terms of the relations between such constants of motion (Column 1 of Table \ref{table:traj}) or in terms of their spatial projections (Column 2 of Table \ref{table:traj}). We summarize these trajectories in Table \ref{table:traj}.  The power distributions of radiation have been calculated for each of these trajectories in \cite{GoodSW:2019} where a point charge moving uniformly accelerated along each path emits constant radiative power.
%\begin{enumerate}[(i)]
    %\item Linear :   $\kappa \neq 0$, $\nu= b =0$;
    %\item Circular : $|\kappa|<|b|$, $\nu=0$; 
    %\item Cusped :   $|\kappa|=|b|$, $\nu=0$;
    %\item Catenary : $|\kappa|>|b|$, $\nu=0$; 
    %\item Helix : $\nu \neq 0$.
%\end{enumerate}

\begin{table}[t!]
\caption{Stationary accelerated trajectories.}
% title of Table
\centering
% used for centering table
\scalebox{0.94}{\begin{tabular}{c c c c}
% centered columns (4 columns)
\hline\hline
%inserts double horizontal lines
Worldline & Spatial projection  & KVF generators  &  Constants of motion \\ [0.5ex]
% inserts table
%heading
\hline
% inserts single horizontal line
Nulltor & Linear & Boost & $\kappa \neq 0$, $b = \nu=0$ \\
% inserting body of the table
Ultrator & Circular & Timelike translation \& rotation & $ |b|>|\kappa | > 0$, $\nu=0$ \\
Parator & Cusped & Timelike translation \& null rotation & $|\kappa|=|b| \neq 0$, $\nu=0$ \\
Infrator & Catenary & Boost \& spatial translation & $|\kappa|>|b| > 0$, $\nu=0$ \\
Hypertor & Helix & Boost \& rotation & $\kappa \neq 0$, $b \neq 0$, $\nu \neq 0$ \\ [1ex]
% [1ex] adds vertical space
\hline
%inserts single line
\end{tabular}}
\label{table:traj}
% is used to refer this table in the text
\end{table}

From a theoretical point of view, the importance of uniform acceleration can hardly be overemphasized in field theory. In addition to the Unruh effect, it is relevant especially in the context of gravitational studies involving the equivalence principle or investigations involving external acceleration effects like moving boundary conditions (see, e.g., for moving mirrors \cite{Davies:1976hi,paper3,GLW,pagefulling,paper1,MPA,paper2}) for the dynamical Casimir effect (for a recent review see  \cite{Dodonov:2020eto}), for the modeling of the onset of Hawking radiation \cite{hawking} during black-hole formation \cite{Juarez-Aubry:2014jba} or cosmological particle creation during expansion \cite{Lib}.

The primary purpose of this paper is to analyze the temperatures recorded along of all the non-trivial stationary worldlines \cite{LetawJ, Rosu:1999ad, Rosu:2000}. As we have discussed above, already for the circular acceleration or Ultrator motion, the temperature recorded will generally depend on frequency. In particular, for an Unruh-DeWitt detector, with which we will be concerned here, the temperature will be a function of the detector transition frequency difference. As we shall see, this frequency dependence is also a feature along the Nulltor (cusped), Infrator (catenary) and Hypertor (helix) motions (see Table \ref{table:traj}). Rosu \cite{Rosu:2005iu} notes that this terminology emphasizing the torsion was first used by Letaw.  We must place special emphasis at this point on the fact that the temperatures for the latter two trajectories have not been discussed much in the literature.

To summarize, the question that we address here is: \\
\\
\textit{What temperatures are recorded along uniformly accelerated motions in flat spacetime? }
\\
\\
We answer this question by a combination of analytic and numerical techniques. We consider that understanding the properties of the temperatures recorded along these trajectories may contribute  an initial step towards a more complete understanding of what might be expected in experimental setups designed to measure the various thermal Unruh-like phenomena. 

As we have mentioned above, and can see from Table \ref{table:traj}, stationary worldlines in Minkowski spacetime have geometric properties independent of proper time, characterized by their invariants \cite{Letaw:1980yv} of curvature, $\kappa$, torsion, $b$, and hypertorsion, $\nu$. In this sense, the temperatures along the stationary worldlines classified by Letaw \cite{Letaw:1980yv} are functions of these invariants. For example, the Unruh temperature, $T_{\rm U}$, is a function of the worldline curvature or proper acceleration, $\kappa$. In view of Table \ref{table:traj}, the temperatures along the Ultrator (circular), Parator (cusped) and Infrator (catenary) worldlines will depend (in addition to the frequency $\omega$) on the curvature, $\kappa$, and torsion, $b$, of the curve traced by the trajectory in Minkowski spacetime. We refer to them respectively as $T_{\rm cir}$, $T_{\rm cus}$ and $T_{\rm cat}$, making reference to their spatial projections. The Hypertor (helix) temperature, $T_{\rm hel}$, will depend (in addition to the frequency $\omega$) on curvature, $\kappa$, torsion, $b$, and hypertorsion $\nu$. We will omit the explicit dependence of the temperatures on $\kappa$, $b$ and $\nu$ for notational reasons, and because this does not lead to confusion in the same way that we do not usually make explicit the dependence of the Unruh temperature on $\kappa$.

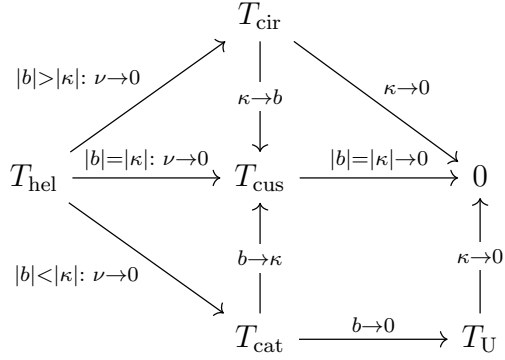
\begin{figure}[!t]
\centering
\begin{tikzcd} 
   &  & T_{\rm cir}\arrow[dd,  "\kappa \to b" description]\arrow[ddrr, "\kappa \to 0"] & &
        \\  \\
   T_{\rm hel} \arrow{uurr}{|b| > |\kappa|: \, \, \nu \to 0} \arrow[rr, "|b| = |\kappa|: \, \,  \nu \to 0"]
   & & T_{\rm cus}\arrow[rr, "|b| = |\kappa| \to 0"] & & 0\\ \\
   &   & T_{\rm cat}\arrow[uu, "b \to \kappa" description]\arrow[leftarrow]{lluu}{|b| < |\kappa|: \, \, \nu \to 0}\arrow[rr, "b \to 0"] & & T_{\rm U}\arrow[uu, "\kappa \to 0" description]
  \end{tikzcd}
  \caption{Limits for the the effective, frequency-dependent temperatures recorded by detectors along stationary curves in Minkowski spacetime.}
  \label{fig:Hierarchy}
  \end{figure}

From an analytic point of view, we shall prove that the limits of Fig. \ref{fig:Hierarchy} hold. The limits presented in this figure have direct physical interpretations, as we shall see in Sec. \ref{Sec:Hierarchy} below. In terms of the minimum speeds, $v_{\rm min}$, along the stationary trajectories; they can be seen as rest-frame low- or high-speed limits. The high-speed limits can be understood in terms of appropriate reference frames connected to the rest frame by Poincar\'e transformations that depend on the parameters $\kappa$, $b$ and $\nu$ (that may diverge in a controlled fashion, see Sec. \ref{Sec:Hierarchy} for details), which connect the different trajectories (at fixed proper time), cf. Fig. \ref{fig:Hierarchy}. We should comment that in Fig. \ref{fig:Hierarchy} there is no $T_{\textrm{cat}} \rightarrow 0$ limit. This is so because for the Infrator or catenary motion it is required that $|b|<|\kappa|$,  so $\kappa \rightarrow 0$ cannot be strictly approached with $b\neq 0$.

Controlling these limits is important for prospective experimental design (see e.g. the discussions in \cite{Rosu:1999ad, Rosu:2000,Rosu:2005iu} for some experimental perspectives). For example, as we have mentioned above, the circular motion has received experimental attention recently \cite{BillTalk, Gooding:2020}. We should mention that the helix motion also has good potential for being studied experimentally, since helix motion can be confined to a ``small" lab region in the near-circular regime, although in general the size of the lab will grow with the time-scale of the experiment due to the helical drift of the motion. From our analytic studies (see Fig. \ref{fig:Hierarchy}), the results of any experiment on helix motion would be expected to asymptotically approach those of circular motion as the hypertorsion vanishes, in the $|b| > |\kappa|$ regime. It could also turn out to be that performing an experiment along helix motion is technically simpler than along circular motion, depending on the experimental details, although this is purely speculative until a concrete experiment is proposed. One could also take the attitude that circular motion is an idealization and in reality the motion of a detector will have a negligible drift, and thus actually follow a helix motion with negligible $\nu$.

From a numerical viewpoint, the asymptotics presented in Fig. \ref{fig:Hierarchy} are also verified numerically in our work, and we have provided plots sampling the temperatures along the stationary worldlines at different values of combinations of the invariant parameters $\kappa$, $b$ and $\nu$. The numerics that we present here also allow us to study the qualitative behavior of these worldline temperatures, since only in the Nulltor and Parator cases are analytic results attainable.  These results allow us to extract information from such curves. 

In particular, we provide numerical evidence that for Parator, Ultrator, Infrator, and Hypertor motions, the effective temperatures dip under the Unruh temperature for small values of $|\omega|$ and go above the Unruh temperature at large $|\omega|$. We obtain in several numerical examples the coldest temperature along each curve, which are found at $\omega = 0$. Of course, from an experimental point of view, the {\emph{hottest}} temperatures are more relevant (at large $|\omega|$), but extracting them from numerical analysis requires a study in its own right, and we think that anyway this deserves attention in an independent article, so we leave it for future work.

A few articles where stationary trajectories other than linear uniform acceleration (Nulltor) have been previously studied in the context of quantum field theory, detector responses, asymptotic states and effective recorded temperatures. (Detector responses, their asymptotic states and effective temperatures are intimately related to the temperature measured by detectors via the generalized, frequency-dependent detailed balance condition \cite{BD}). These works necessarily have some overlap with the results that we present in this paper, especially in the Parator (cusped) and Ultrator (circular) cases.

In \cite[Sec. 5.2]{Louko:2006zv}, Louko and Satz have made some investigations of the transition rates of detectors following stationary worldline  trajectories by the corresponding Killing vector,  classified by Letaw  \cite{Letaw:1980yv} and reviewed by Rosu \cite{Rosu:2000,Rosu:2005iu}, which we show on Table \ref{table:traj}. In addition to studying the Unruh effect, Louko and Satz have given a closed formula for the transition rate of a detector following the Parator (cusped) motion, from which an effective temperature can be inferred. Such effective temperature has been studied by two of us in \cite{BD}, and was essentially already given in \cite{Letaw:1980yv}. The detector responses or effective temperatures for the remaining trajectories -- Ultrator, Infrator or Hypertor -- have not been studied in detail in \cite{Louko:2006zv}.

As we have mentioned, the circular motion has been studied, in addition to \cite{Leinaas.Bell,Bell:Leinaas}, by Unruh in \cite{Unruh:circ} and in \cite{Sriramkumar:1999nw, Sriram}.  A thorough study of the circular motion in Minkowski spacetime will appear in \cite{Gooding:2020}.\footnote{We thank Jorma Louko for bringing this work in preparation to our attention.} We should also mention that circular geodesics have been studied in curved spacetimes in \cite{Hodgkinson:2012mr, Hodgkinson:2014iua, Ng:2014kha} in $2+1$ and $3+1$ dimensions. See also \cite{Stargen:2017xii} for a study of a detector following circular motion interacting with a polymer-quantized field and \cite{Gutti:2010nv, Louko:2017emx} in the context of modified dispersion relations.

To the best of our knowledge, the asymptotic states of detectors following the Infrator (catenary) and Hypertor (helix) motions and the effective temperatures that they register have not been thoroughly studied in the literature. One of the purposes of this paper is to fill in this literature gap.

%In such  trajectories  ($\Delta x^2$) depends on the proper time difference between the two points. As a result, a detector's transition rate in the infinite past is independent of proper time.   Stationary worldlines are the trajectories of timelike Killing vectors in flat space, where an independent definition of ‘particles’ exists via the positive and negative frequency decomposition (see \cite{Louko:2006zv} and references therein).  

The paper is organized as follows. In Sec. \ref{thermalize}, we briefly describe the Unruh-DeWitt detector model that we study in this paper, and summarize the main result obtained by two of us in \cite{BD}. Namely, that detectors following stationary trajectories in Minkowski spacetime reach an equilibrium state at late times at an effective temperature that can be obtained from a generalized detailed balance condition. This justifies in which sense these frequency-dependent, effective temperatures deserve to be called temperatures in the first place. Sec. \ref{Sec:Hierarchy} contains the main results of this work, including the numerical analysis that we have discussed above and a proof of each one of the limits appearing in Fig. \eqref{fig:Hierarchy}, as well as a discussion on their physical meanings. Finally, Sec. \ref{conclusions} contains the final remarks and future perspectives of this work.  

Spacetime points are denoted by Roman characters $(\x)$. We henceforth set $\hbar = c = k_B = 1$ in the understanding that the physical temperatures can be restored using Eq. \eqref{UnruhPhys}.

\section{Thermalization of stationary detectors}\label{thermalize}

 We consider an Unruh-DeWitt detector  \cite{Unruh:1976db,Dewitt:1979} moving along a trajectory $\x(\tau)$ in four-dimensional Minkowski spacetime, with $\tau$ the proper time of the detector. We model the detector by a two-level system (qubit) with transition frequency difference $\omega \in \mathbb{R}$ interacting with a massless scalar quantum field in the Minkowski vacuum state. If $\omega > 0$ the detector is excited, while if $\omega < 0$ the detector is deexcited. The interaction Hamiltonian is 
\begin{equation}
\hat {\rm H}_{\text{int}}(\tau)=c\chi(\tau)\hat \mu(\tau)\otimes\hat \Phi\left(\x(\tau)\right),
\label{Hint}
\end{equation}
where $c$ is small a coupling constant and $\chi\in C_0^{\infty}(\mathbb{R})$ is a real smooth compactly supported function, which specifies how the interaction is switched on and off between the detector's monopole moment operator $\hat{\mu}(\tau)$ and the field pulled back to the detector's worldline $\hat \Phi\left(\x(\tau)\right)$.

We are concerned with the cases in which the detector follows a stationary trajectory \cite{Letaw:1980yv}, i.e., when the trajectory $\x(\tau)$ is along the integral curves of a Killing vector that is timelike in a neighbourhood of the trajectory. In these cases, the Wightman two-point correlation function of the field will depend only on the proper time interval, $\tau-\tau'$, between the two worldline points, when pulled back along the detector's trajectory. In our case, the pullback of the Wightman function is given by (see, e.g., \cite{Birrell})
\begin{align}
G^+(\x(\tau), \x(\tau'))=-\frac{1}{4\pi^2} \frac{1}{(t(\tau)-t(\tau')-i\epsilon)^2-|\mathbf{x}(\tau)-\mathbf{x}(\tau')|^2},
\label{wightman}
\end{align}
with $\epsilon\to 0^+$ in the distributional sense. Owing to the discussion above, we shall henceforth denote the pullback by $\mathcal{W}^+(\tau-\tau') := G^+(\x(\tau), \x(\tau'))$.

For such stationary trajectories, if the generalized detailed balanced condition for the pullback of the Wightman function of the field along the detector worldline, $\widehat{\mathcal{W}}^+(-\omega)=e^{\beta(\omega)\omega}\widehat{\mathcal{W}}^+(\omega)$, is satisfied, when the switching function, $\chi$, is stretched adiabatically for a long time period, it has been discussed in \cite{BD}  that the  detector will reach an equilibrium state (at leading order in the coupling $c$) 
\begin{align}
    \hat \rho
       = \frac{1}{1 + \expe^{-\omega/T_{\rm eff}(\omega)}} \begin{pmatrix} 
       1 & 0\\ 0 & \expe^{-\omega/T_{\rm eff}(\omega)} \\
       \end{pmatrix},
 \label{AsymptState}     
\end{align}
at an effective, frequency dependent temperature
\begin{align} 
T^{-1}_{\rm eff}(\omega) =  \frac{1}{\omega} \ln \left(\frac{\widehat{\mathcal{W}}^+(-\omega)}{\widehat{\mathcal{W}}^+(\omega)}\right),\label{temp}
\end{align} where we have denoted by
\begin{equation}
\widehat{\mathcal{W}}^+(\omega)=\int_{-\infty}^{+\infty}ds\, \expe^{-i\omega s} \mathcal{W}^+(s)
\end{equation}
the Fourier transform of the pullback of the Wightman function. 

The purpose of Sec. \ref{Sec:Hierarchy} will be to study in detail $T_{\rm eff}$ in the case of the Nulltor (linear), Parator (cusped), Ultrator (circular), Infrator (catenary) and Hypertor (helix) trajectories.

\section{The effective temperatures}\label{Sec:Hierarchy}

In this section, we use a combination of analytic and numerical techniques to study the effective temperatures of the non-trivial stationary trajectories in Minkowski spacetime described in Table \ref{table:traj}.

On the analytic side, we will show each of the limits appearing in Fig. \ref{fig:Hierarchy}. The strategy will be to express the effective temperatures along each of the trajectories in terms of integrals -- the Fourier transform of the pullback of the Wightman function along the trajectory -- using Eq. \eqref{temp}, and to understand that the relevant limits can be obtained in each case using dominated convergence. Thus, in this section we show that the limits appearing in Fig. \ref{fig:Hierarchy} are all rigorous.

As we shall see, the limits appearing in Fig. \ref{fig:Hierarchy} are intimately related to physical asymptotic regimes, such as high or low speeds along the worldline.

On the numerical side, we shall explore the space of parameters of $\kappa$, $b$, and $\nu$ (curvature, torsion and hypertorsion respectively) for the stationary worldlines of Table \ref{table:traj} and verify our analytic limits numerically, as well as explore the behavior of the effective temperatures as a function of frequency for different values of $\kappa$, $b$, and $\nu$. As we have discussed in the Introduction, we shall verify numerically that generically the effective temperatures of Parator, Ultrator, Infrator and Hypertor can dip under the Unruh temperature at low frequencies and go above the Unruh temperature at high frequencies.

This section contains necessarily some well-known results. Namely, the temperatures along the Nulltor (linear) and Parator (cusped) motions are known in closed form. In the Nulltor case this is the Unruh temperature along a linearly uniformly accelerated trajectory, see, e.g., \cite{Unruh:1976db, BD, FBL, Louko:2006zv,  Birrell, DMCA, DeBievre }. For the Parator trajectory, see, e.g., \cite{BD, GoodSW:2019, Louko:2006zv,  Proceedings}. The effective temperature in the Ultrator (circular) case was studied already in \cite{Leinaas.Bell, Bell:Leinaas, Unruh:circ}, and is also discussed thoroughly in the recent work \cite{Gooding:2020}, but the rigorous analytic results as $b \to \kappa$ and as $\kappa \to 0$ that we present here are new. 

To the best of our knowledge, no previous thorough studies of the Infrator (catenary) and Hypertor (helix) cases exist.

%The response of detectors following some stationary trajectories has been studied in previous literature. In particular, this has been the case for the Nulltor (linear), Ultrator (circular) and Parator (cusped) motions. The asymptotic state of the detectors following these trajectories at late times deduced are given by Eq.~ \eqref{AsymptState} with temperatures \eqref{temp}.

\subsection{Nulltor (linear) worldline}

The temperature for Nulltor motion is the Unruh temperature $T_{\rm U} = \kappa/2\pi$, at which a uniformly linear accelerated detector following the worldline
\begin{align}
{\x}(\tau)=\left(\kappa^{-1}\sinh(\kappa\tau), \, \kappa^{-1}\cosh(\kappa\tau),\, 0, \, 0\right),
\label{NulltorMotion}
\end{align}
thermalizes after a long interaction time with the field (for a precise meaning of long interaction time see, e.g., \cite{BD, FBL, DMCA, DeBievre, DM}). The trajectory Eq.~\eqref{NulltorMotion} depends only on the curvature $\kappa$, which is the acceleration along the motion, while torsion and hypertorsion are identically zero. It is obvious that as $\kappa \to 0$, i.e., as the acceleration vanishes, $T_{\rm U} \to 0$.

\subsection{Parator (cusped) worldline} 
\label{sec:cusp}

\begin{figure}[t!]
\begin{center}
{\rotatebox{0}{\includegraphics[width=5.0in]{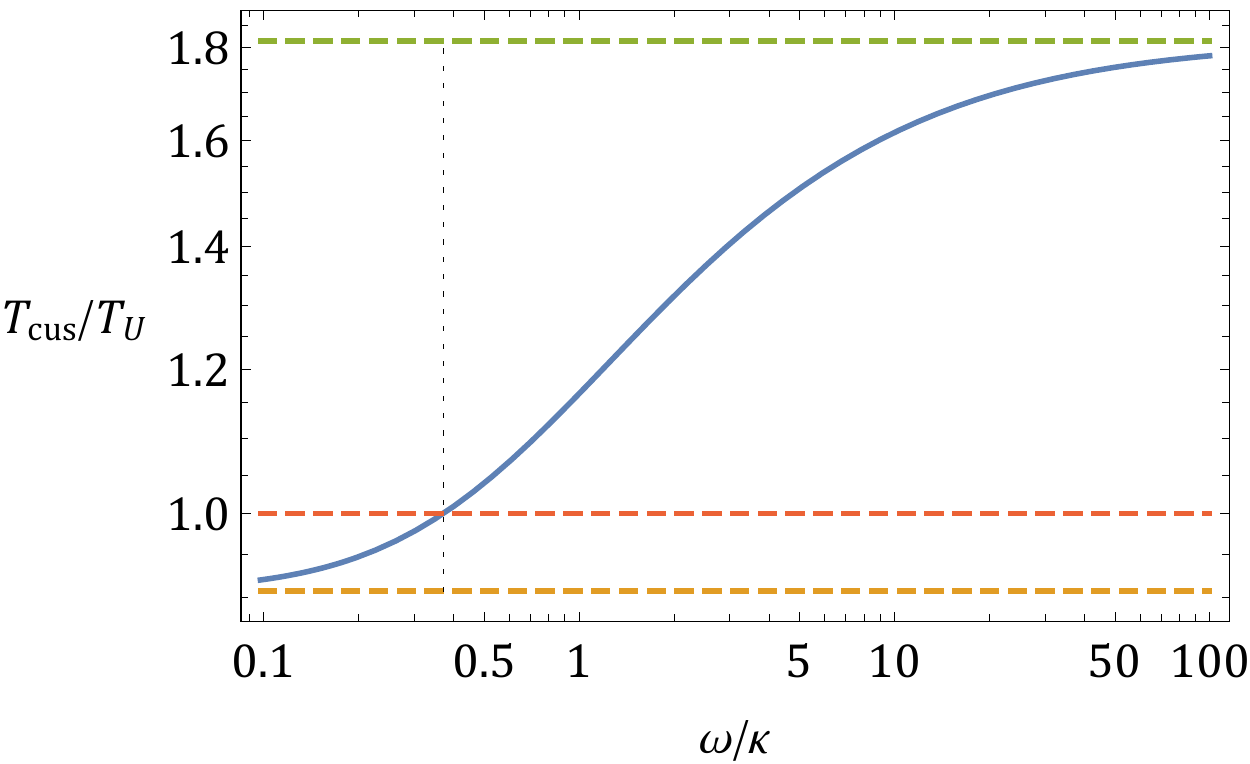}}} 
\caption{\label{fig:cusp} Effective temperature along the Parator worldline. The coldest temperature $T_{\rm cus}(0)/T_{\rm U} = \pi/(2\sqrt{3})$  is indicated by a yellow dotted line. The hottest temperature $T_{\rm cus}(\infty)/T_{\rm U} = \pi/\sqrt{3}$ is indicated by a green dotted line.  $T_{\rm cus} = T_{\rm U}$  at characteristic frequency $\omega/\kappa \approx 0.372$. } 
\end{center}
\end{figure}  

%The cusp, first found in \cite{Letaw:1980yv}, is a particularly unique motion among the five non-trivial trajectories due to its analytical tractability (despite this only a few papers have investigated it, see e.g. \cite{BD, GoodSW:2019, Proceedings}).  

The cusped motion has non-vanishing curvature and torsion, with $\kappa = b$, and is parametrized by proper time as
\begin{align}
\x(\tau)=\left(\tau+\frac{1}{6}\kappa^2\tau^3,\, \frac{1}{2}\kappa\tau^2,\, \frac{1}{6}\kappa^2\tau^3,0\right).
\label{cusp}
\end{align}
The (unregularized) pullback of the Wightman functions to the trajectory is
\begin{align}
\mathcal{W}^{+}_{\rm cus}(s)=-\frac{1}{4\pi^2}\frac{1}{s^2+\frac{\kappa^2}{12}s^4},
\end{align}
where $s = \tau - \tau'$, and the effective, frequency-dependent temperature  can be obtained in closed form by complex-analytic techniques \cite{BD, Letaw:1980yv, Louko:2006zv}, 
\be T^{-1}_{\rm cus}(\omega) = \frac{1}{\omega} \ln \left( \frac{\omega \Theta(\omega)+\frac{\kappa}{2\sqrt{12}}\expe^{-\sqrt{12}|\omega|/\kappa}}{-\omega \Theta(-\omega)+\frac{\kappa}{2\sqrt{12}}\expe^{-\sqrt{12}|\omega|/\kappa}}\right).
\label{Tcus}
\ee

We plot Eq. \eqref{Tcus} in Fig.~\ref{fig:cusp}.  Due to this analytic form, it is easy to find the asymptotics and special values. In particular,
\begin{align}
\lim_{\kappa \to 0} T_{\rm cus}(\omega) = 0.
\end{align}The coldest effective temperature achieved along the Parator trajectory is $T_{\rm cus}(0) = \frac{\kappa}{4 \sqrt{3}} = \frac{\pi}{2\sqrt{3}} T_{\rm U} \approx 0.90 T_{\rm U}$, dipping under the Unruh temperature.  The hottest temperature is twice the lowest temperature achieved, found at $T_{\rm cus}(\infty) = T_{\rm cus}(-\infty) = \frac{\pi}{\sqrt{3}} T_{\rm U}$. The characteristic frequency, where the effective temperature is that of the Unruh temperature, $T_{\rm cus}(\omega) /T_U = 1$, is found by solving the root of the expression:
\be \expe^{2 \pi  \bar{\omega}}-4 \sqrt{3}  \expe^{2 \sqrt{3} \bar{\omega}}\bar{\omega}-1 = 0, \ee
which is approximately $\bar{\omega}:=\omega/\kappa \approx 0.372$. 

\subsection{Ultrator (circular) worldline}
\label{sec:Cir}

\begin{figure}[t!]
\begin{center}
{\rotatebox{0}{\includegraphics[width=5.0in]{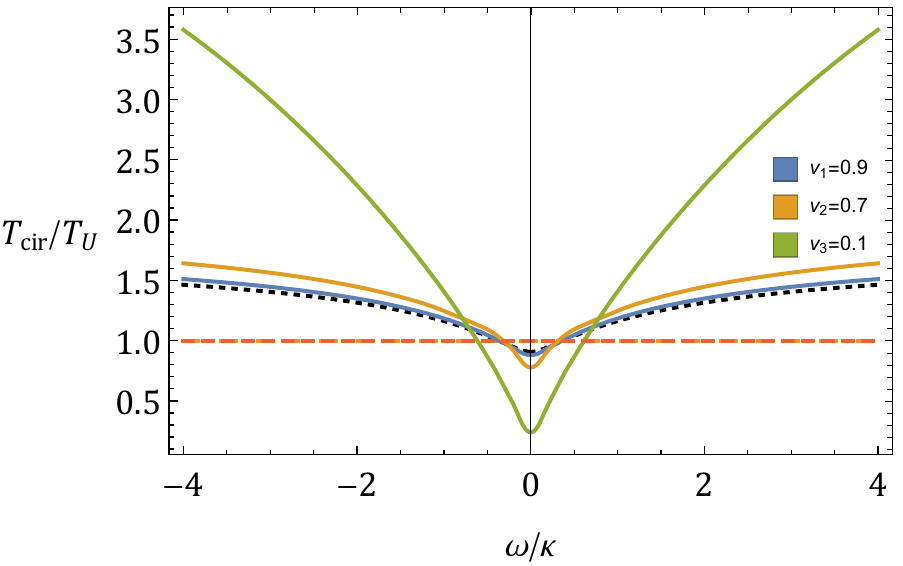}} {\caption{\label{fig:circle} Effective temperature along the Ultrator (circular) worldline. The Unruh temperature, $T_{\rm U}$, is indicated by a red dotted line. The Parator (cusped) motion temperature, $T_{\rm cus}$, is indicated by a black dotted line. At large values of $v$ the Ultrator temperature asymptotes the Parator temperature.}}} 

\end{center}
\end{figure}  

\begin{figure}[t!]
\begin{center}
\includegraphics[scale=0.65]{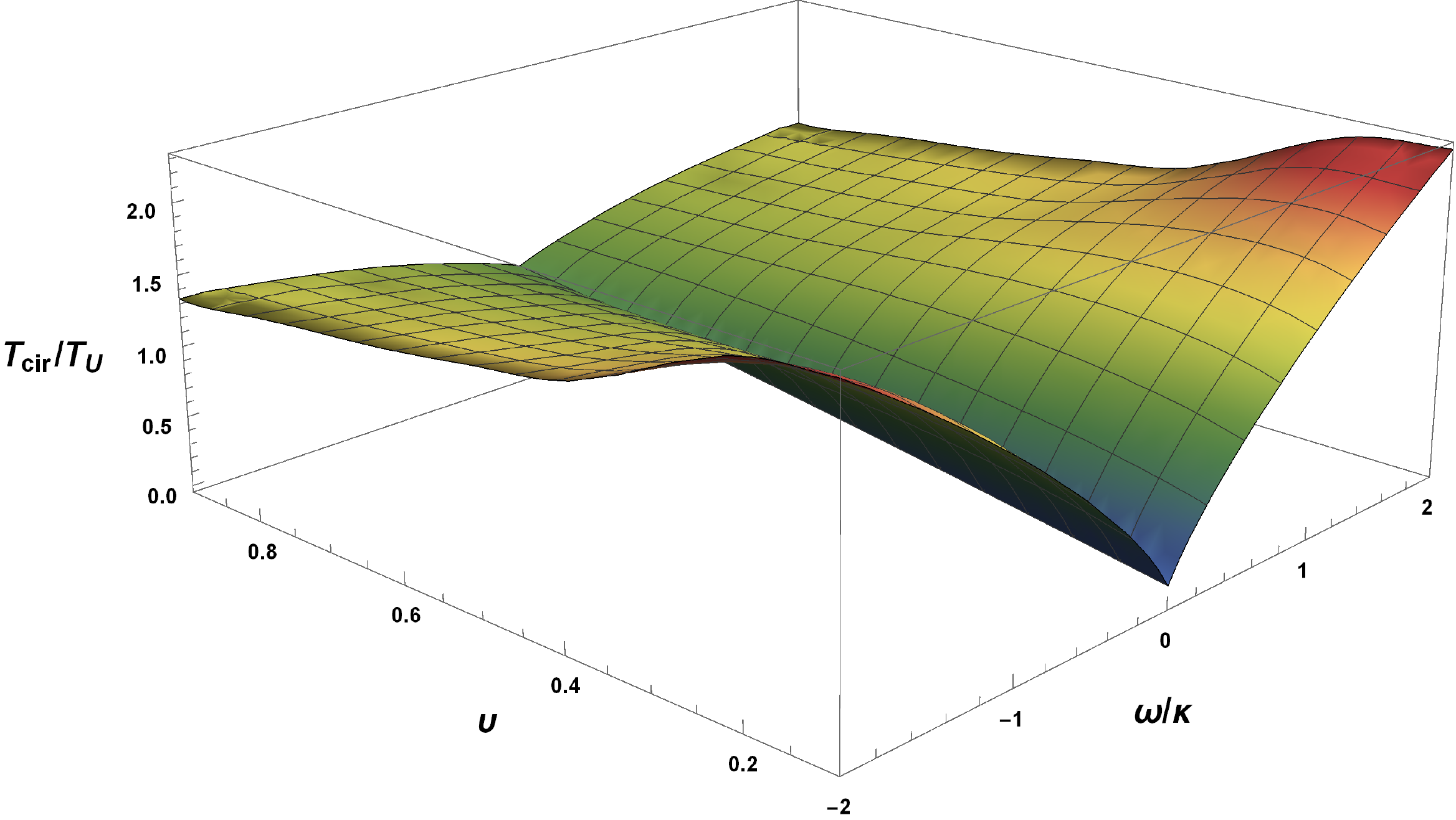}
\caption{The effective temperature of a detector following a circular trajectory scaled to the Unruh temperature as a function of $\omega/\kappa$ and the velocity $0.1\leq\upsilon\leq0.9$. In the limit $\omega/\kappa\to 0$, a slowly moving detector with velocity $\upsilon=0.1$ registers the coldest effective temperature $T_{\rm eff}/T_U \approx 0.099$. The lowest effective temperature obtained by a detector moving with $\upsilon=0.9$ is $T_{\rm eff}/T_U \approx 0.84$.}\label{fig:3Dcircle}
\end{center}
\end{figure}

A circularly moving detector follows the worldline  \cite{Unruh:circ,LePf}
\begin{align}
\x(\tau)=\left(\gamma \tau,\, r\cos(\gamma \Omega \tau),\, r\sin(\gamma \Omega \tau) ,0\right),\label{circtraj}
\end{align}
where  $\Omega$ is the angular velocity of the detector, $r$ is the radius  of the circular orbit, $\gamma=1/\sqrt{1-\upsilon^2}$ is the Lorentz factor and $\upsilon=\Omega r$  is the velocity of the detector. In terms of curvature invariants, $\Omega = b(1-\kappa^2/b^2)$, and $v = \kappa/b$, where $\kappa$ is the proper acceleration and $b$ is the torsion, which satisfy $|\kappa| < |b|$. It has vanishing hypertorsion $\nu = 0$. 

Setting $s = \tau - \tau'$, the pullback of the Wightman function along, Eq. \eqref{wightman}, the circular worldline is
%\begin{align}
%\mathcal{W}^+_{\rm circle}(\tau-\tau')=\frac{1}{4\pi^2} \frac{1}{-(\tau-\tau'-i\epsilon)^2+(x-x')^2+(y-y')^2+(z-z')^2}.
%\label{wightman}
%\end{align}
%Inserting Eq.~(\ref{circtraj}), and setting $s\equiv \tau-\tau'$, we have
\begin{align}
\mathcal{W}^+_{\rm cir}(s)=\frac{1}{4\pi^2} \frac{1}{-\gamma^2 s^2 + 2 r^2(1-\cos{(\gamma \Omega s)}) }.
\label{Wcircular}
\end{align}
As far as we can see, the circular motion effective temperature cannot be obtained analytically, as is the case of the linearly uniformly accelerated and cusped motions, because the integral defining the Fourier transform of Eq.~\eqref{Wcircular} is intractable. In Appendix \ref{App:Cir}, however, we show that the following limits hold
\begin{subequations}
\label{circlimits}
\begin{align}
\lim_{\kappa \to 0} T_{\rm cir}(\omega)  & = 0,  \label{circKappa0}\\
\lim_{b \to \kappa} T_{\rm cir}(\omega) & = T_{\rm cus}(\omega).
\label{circKappab}
\end{align}
\end{subequations}

In other words, as the curvature vanishes, the temperature of the circular motion vanishes. This is the low-speed limit $v \to 0$. 

As $b \to \kappa$, the temperature of circular motion coincides with that of the cusped motion. In the rest frame this means that $v \to 1$, $r \to \infty$ and $\Omega \to 0$. We can see, however, that passing to the frame $t' = \gamma(t- v y)$, $x' = x - r$ and $y' = \gamma (y - v t)$ with a translation in the $x$-direction and a boost in the $(t,y)$-plane, one has that
\begin{subequations}
\label{Circ:Poincare}
\begin{align}
t'(\tau) & = \gamma \left( \gamma \tau - v r \sin(\gamma \Omega \tau) \right) \xrightarrow[b \to \kappa]{} \tau + \frac{\kappa^2 \tau^3}{6}, \\
x'(\tau) & = r \cos(\gamma \Omega \tau) - r \xrightarrow[b \to \kappa]{} - \frac{\kappa \tau^2}{2}, \\
y'(\tau) & = \gamma \left( r \sin(\gamma \Omega \tau) -  v\gamma \tau \right) \xrightarrow[b \to \kappa]{} \frac{\kappa^2 \tau^3}{6},
\end{align}
\end{subequations}
which (upon a reflection in the $x$-axis) gives the cusped trajectory Eq.~\eqref{cusp} in the new frame. Note that the transformations Eq.~\eqref{Circ:Poincare} diverge suitably in the $b \to \kappa$ limit, but in such a way that the image under the transformations are indeed finite in the $b \to \kappa$.

%We must emphasize at this point that we do not mean that at high speeds the circular motion Eq.~\eqref{circtraj} tends to the cusp motion Eq.~\eqref{cusp}, but rather that at high speeds the temperature registered along circular motion will coincide with the temperature registered along cusp motion, cf. Eq. \eqref{circKappab}.

%In the limit of high-speeds the circular trajectory has temperature that approaches the cusp temperature, which we will see in the following section.  The circular temperature function cannot be obtained analytically because the integration is intractable, and so we resort to an approximation. For a high-speed motion approximation, we consider the cosine function in the denominator of the Wightman function and expand it to fourth order in proper time, 
%\begin{align}
%\cos{(\gamma \Omega s)}=1- \frac{1}{2}\gamma^2 \Omega^2 s^2+ %\frac{1}{24}\gamma^4 \Omega^4 s^4.
%\end{align}

%We find after taking the Fourier transform, the effective temperature gives that the coldest temperature of Ultrator at high-speeds as 0.907 and the hottest temperature at 1.81 which coincides with the limits of the cusp motion in Fig.~(\ref{fig:cusp}).  We have obtained for a moving detector at velocity $\upsilon=0.1$, the coldest temperature $T_{\rm eff}/T_U \simeq 0.099$ in the degenerate limit $\omega\to0$. 

Temperatures along circular motion for a selection of values of $v$ are presented in Fig.~\ref{fig:circle}, where at large values of $v$ the numerical results can be seen to converge to the cusped motion curve (dotted). Fig.~\ref{fig:3Dcircle} further shows numerical solutions as a function of velocity, $v$.

\subsection{Infrator (catenary) worldline}
\label{Sec:Infrator}

\begin{figure}[t!]
\begin{center}
{\rotatebox{0}{\includegraphics[width=5.0in]{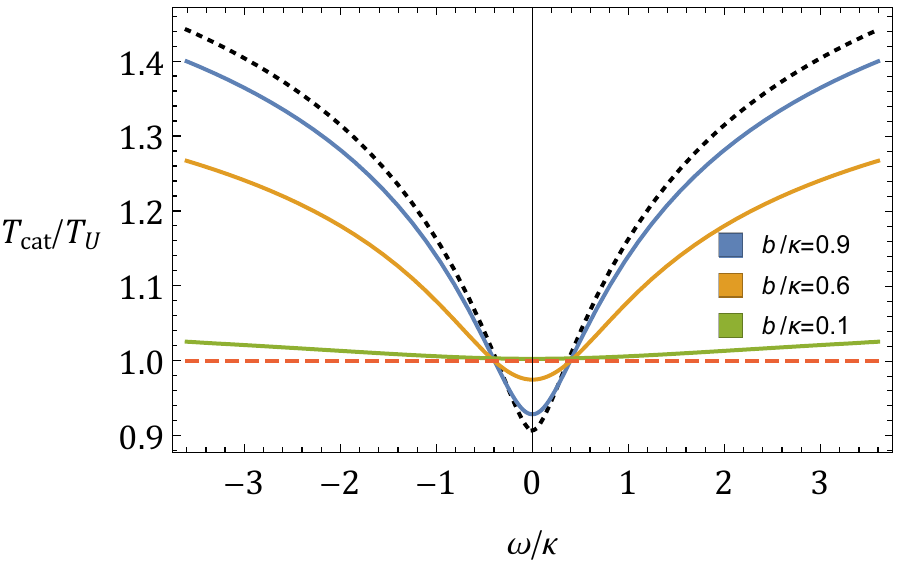}}} 
{\caption{\label{fig:WL_Cat} Effective temperature along the Infrator (catenary) worldline. The Unruh temperature, $T_{\rm U}$, is indicated by a red dotted line. The Parator (cusped) motion temperature, $T_{\rm cus}$, is indicated by a black dotted line. At large values of $v_{\rm min}$ the Infrator temperature asymptotes the Parator temperature. As small values of $v_{\rm min}$ the Infrator temperature asymptotes the Unruh temperature. }
%The effective temperature for the catenary worldline scaled to the angular velocity  of the detector  for three different values of the velocity (Infrator-Catenary).\maksat{Where black dotted line demonstrates the  temperature limit of Infrator motion which tend to Paratore motion as $b \to \kappa$. Likewise, red dashed line shows the  the Unruh temperature  and  black dashed line illustrates the coldest effective temperature.
}    
\end{center}
\end{figure} 

\begin{figure}[t!]
\begin{center}
\includegraphics[scale=0.65]{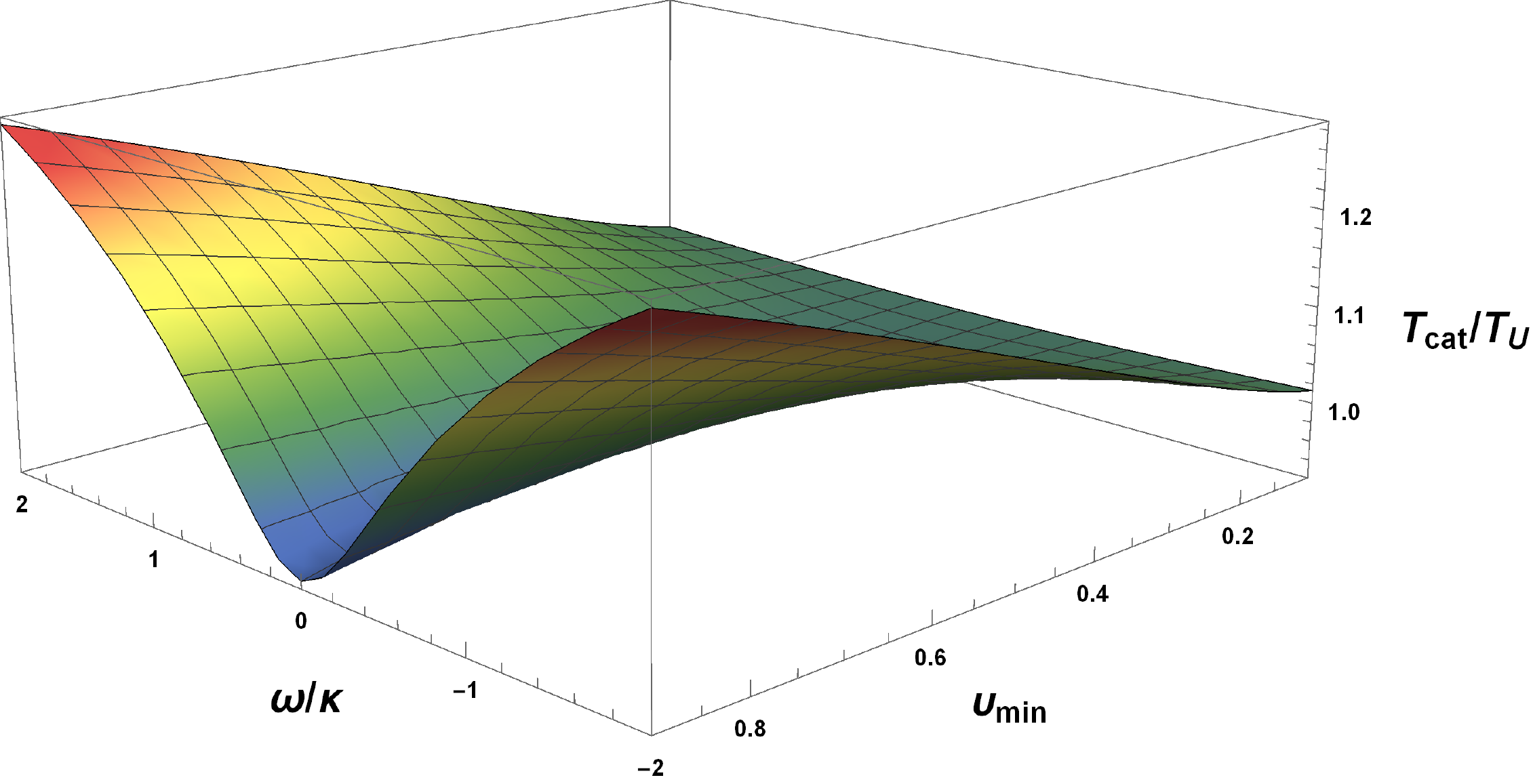}
\caption{ \label{fig:3Dcat} The effective temperature of a detector following a catenary trajectory scaled to the Unruh temperature as a function of $\omega/\kappa$ and the minimum velocity $0.1\leq\upsilon_{min}\leq0.9$.
In the limit $\omega/\kappa\to0$, a detector moving with velocity $\upsilon_{\rm min} =0.9$  reaches the coldest temperature $T_{\rm eff}/T_U \approx 0.92$. A slowly moving detector with velocity $\upsilon=0.1$ reaches the Unruh temperature $T_{\rm eff}/T_U \approx 1$.
}
\end{center}
\end{figure}

As shown in Table \ref{table:traj}, the Infrator motion is generated by a combination of the boost and spatial translation Killing vector fields of Minkowski spacetime. It has vanishing hypertorsion, $\nu = 0$, and satisfies the relation $ |\kappa | > |b|$ between curvature (acceleration) and torsion, leading to no wrap-around like in circular motion. The spatial projection is instead the familiar catenary, but the speed changes along the trajectory. The worldline can be parametrized in terms of proper time as
\begin{align}
\x(\tau)=\sigma^{-2} \left(\kappa \sinh{(\tau \sigma)},\, \kappa \cosh{(\tau \sigma)},\, 0 ,\tau b \sigma \right),\label{cattraj}
\end{align}where $\sigma^2 := \kappa^2 - b^2$ and $v_{\rm min} = b/\kappa$ is the minimum speed of the detector.  The detector will always maintain a speed higher than this minimum, unlike in the Ultrator case, where speed remains constant. The pullback of the Wightman function, Eq.~ \eqref{wightman}, to the curve defined by Eq.~\eqref{cattraj} is 
\begin{align}
\mathcal{W}^+_{\rm cat}(s) = \frac{\sigma^4}{4 \pi ^2 \left(\sigma^2 \left(b^2 s^2+2\right)-2 \left(\sigma^2+b^2\right) \cosh (\sigma s)+2 b^2\right)},
\label{W+Infrator}
%\mathcal{W}^{+}_{\rm cat}(\tau-\tau')=\frac{1}{4\pi^2} \frac{\sigma^2}{2\kappa^2(1-\cosh{\sigma(\tau-\tau'))+\epsilon^2\sigma^4 +b^2\sigma^2(\tau-\tau')^2+\mathcal{C}_{\epsilon}} },
\end{align}where we have omitted the $\epsilon$ regulator.
%where
%\begin{align}
%\mathcal{C}_{\epsilon}\equiv 4i\epsilon \kappa \sigma^2\sinh{\frac{\sigma}{2}(\tau-\tau')}\cosh{\frac{\sigma}{2}(\tau+\tau')}.
%\end{align}
%By using $s\equiv \tau-\tau'$ and assuming that $\epsilon^+ \to 0$, we have
%\begin{align}
%\mathcal{W}^{+}_{\rm cat}(s)=\frac{1}{4\pi^2} \frac{\sigma^2}{2\kappa^2(1-\cosh{(\sigma s)})+\epsilon^2\sigma^4 +b^2\sigma^2 s^2+4i\epsilon \kappa \sigma^2\sinh{(\frac{\sigma}{2}s)}\cosh{(\frac{\sigma}{2}s)} },
%\end{align}
%and utilizing a half-angle formula on $\mathcal{C}_\epsilon$, where $s$ can be $\tau +\tau'$ for small $\epsilon^+$,  gives
%\begin{align}
%\mathcal{W}^{+}_{\rm cat}(s)=\frac{1}{4\pi^2} \frac{\sigma^2}{2\kappa^2(1-\cosh{(\sigma s)})+\epsilon^2\sigma^4 +b^2\sigma^2 s^2+2i\epsilon \kappa \sigma^2\sinh{(\sigma s)}}\label{Wcat}.
%\end{align}

From Eq.~\eqref{W+Infrator}, using Eq.~\eqref{temp} we can examine the temperature detected by an Unruh-DeWitt detector following the Infrator motion, and infer the asymptotic state of the detector from Eq.~\eqref{AsymptState} \cite{BD}.

We show in Appendix \ref{App:Cat}, by a dominated convergence argument, that the following limits hold for the catenary motion effective temperature:

\begin{subequations}
\begin{align}
\lim_{b \to 0} T_{\rm cat}(\omega) &  =  \frac{\kappa}{2 \pi},  \\
\lim_{b \to \kappa} T_{\rm cat}(\omega) &  =  T_{\rm cus}(\omega) .
\end{align}
\end{subequations}

In other words, as the torsion vanishes, the effective temperature approaches the Unruh temperature, while as the torsion approaches the curvature, the effective temperature of the catenary motion coincides with that of cusped motion.

Physically, our asymptotic limits indicate that for low minimum speeds, $v_{\rm min}$ ( $b \to 0$), one recovers the Unruh temperature.

At high minimum speeds in the rest frame, $v_{\rm min}$ ($b \to \kappa$  or $\sigma \to 0$), the temperature coincides with that of the cusped motion, reported in Sec. \ref{sec:cusp}. We can see that passing to the frame $t' = \gamma(t - v_{\rm min} z)$, $x' = x - \frac{\kappa}{\sigma^2}$, and $z = \gamma(z - v_{\rm min} t)$ with a translation in the $x$-direction and a boost in the $(t,z)$-plane, one has that
\begin{subequations}
\begin{align}
t'(\tau) & = \gamma \left( \frac{\kappa}{\sigma^2} \sinh(\tau \sigma) - \frac{\tau b^2}{\kappa \sigma}  \right) \xrightarrow[b \to \kappa]{} \tau + \frac{\kappa^2 \tau^3}{6}, \\
x'(\tau) & = \frac{\kappa}{\sigma^2} \cosh(\tau \sigma) - \frac{\kappa}{\sigma^2} \xrightarrow[b \to \kappa]{}  \frac{\kappa \tau^2}{2}, \\
z'(\tau) & = \gamma \left( \frac{\tau b}{\sigma} - \frac{b}{\sigma^2} \sinh(\tau \sigma) \right) \xrightarrow[b \to \kappa]{} -\frac{\kappa^2 \tau^3}{6},
\end{align}
\end{subequations}
which (upon a reflection in the $z$-axis) gives the cusped trajectory Eq.~\eqref{cusp} in the new frame.

The effective, frequency-dependent temperatures interpolating for $0 < b < \kappa$ are obtained numerically. The numerical results are reported in Fig.~\ref{fig:WL_Cat} and Fig.~\ref{fig:3Dcat}.

\subsection{Hypertor (helix) worldline}
\label{Sec:Hypertor}

The hypertor trajectory is a superposition of both circular and linear accelerated motions and is fully $(3+1)$-dimensional.  It includes a non-vanishing hypertorsion, which is a parameter on the other plane of proper angular velocity.  %Its power distribution was calculated previously \cite{GoodSW:2019}.  
The parametric equation of the hypertorsional motion  is written as
\begin{align}
\x(\tau)=\left(\frac{\alpha}{R_+}\sinh{R_+ \tau},\, \frac{\alpha}{R_+}\cosh{R_+ \tau},\, \frac{\beta}{R_-}\cos{R_- \tau} ,\frac{\beta}{R_-}\sin{R_- \tau}\right),\label{hyptraj}
\end{align}
where $\alpha :=\frac{\Delta}{R}$ and $\beta :=\frac{\kappa b}{R \Delta}$ and
\begin{align}  
\Delta^2 & := \frac{1}{2}( R^2 +\kappa^2 + b^2 + \nu^2 ), &  \\
 R^2  & := R_+^2 + R_-^2,  &  R_{\pm}^2 & := {\sqrt{A^2+B^2} \pm A} ,
 \\A & :=\frac{1}{2} \left(\kappa^2-\nu ^2-b ^2\right), &  B & := \kappa  \nu .
 \label{Helix-Parameters}
 \end{align}
All coefficients in this motion are a combination of the curvature or proper acceleration, $\kappa$, torsion, $b$, and hypertorsion, $\nu$. 

The minimum velocity for the motion is
\be
 v_{\textrm{min}} = \frac{\kappa b}{\Delta^2}= \frac{\beta}{\alpha}. \ee
A particle moving along the Hypertor trajectory will always maintain a speed above the minimum velocity, $v_{\textrm{min}}$.  The minimum velocity of the trajectory obeys the following equation in terms of the motion invariants
\be
 \nu = \pm\sqrt{\kappa b\left(\frac{1+v_{\textrm{min}}^2}{v_{\textrm{min}}}\right)-(\kappa^2+b^2)}. 
 \label{Hypertorvmin}
 \ee
Physically, it can be verified from Eq. \eqref{Hypertorvmin} that the higher the minimum speed, the less hypertorsion and conversely the lower the speed, the higher hypertorsion.

The pullback of the Wightman function along the Hypertor motion is
%\begin{align}
%\mathcal{W}^{+}_{\rm helix}(s)=\frac{1}{4 \pi ^2}\frac{1}{\frac{2 \alpha ^2 }{R_+^2}\left(1-\cosh \left(R_+ s\right)\right)+\frac{2 \beta ^2 }{R_-^2}\left(1-\cos
  % \left(R_- s\right)\right)+\frac{2 i \alpha  \epsilon }{R_+} \sinh \left(R_+ s\right)+\epsilon ^2},
%\end{align}
\begin{align}
\mathcal{W}^{+}_{\rm hel}(s)=\frac{1}{4 \pi ^2}\frac{1}{\frac{2 \alpha ^2 }{R_+^2}\left(1-\cosh \left(R_+ s\right)\right)+\frac{2 \beta ^2 }{R_-^2}\left(1-\cos
 \left(R_- s\right)\right)}.
\end{align}In view of Table \ref{table:traj}, we can consider three cases for the invariants of motion
\begin{enumerate}
\item $0 < |\kappa| < |b|$ , $\nu \neq 0$,
\item $|\kappa| = |b| \neq 0$ , $\nu \neq 0$, and
\item $|\kappa| >| b| > 0$ , $\nu \neq 0$ .
\end{enumerate}

\begin{figure}[t!]
    \centering
    \subfigure{\includegraphics[width=0.45\textwidth]{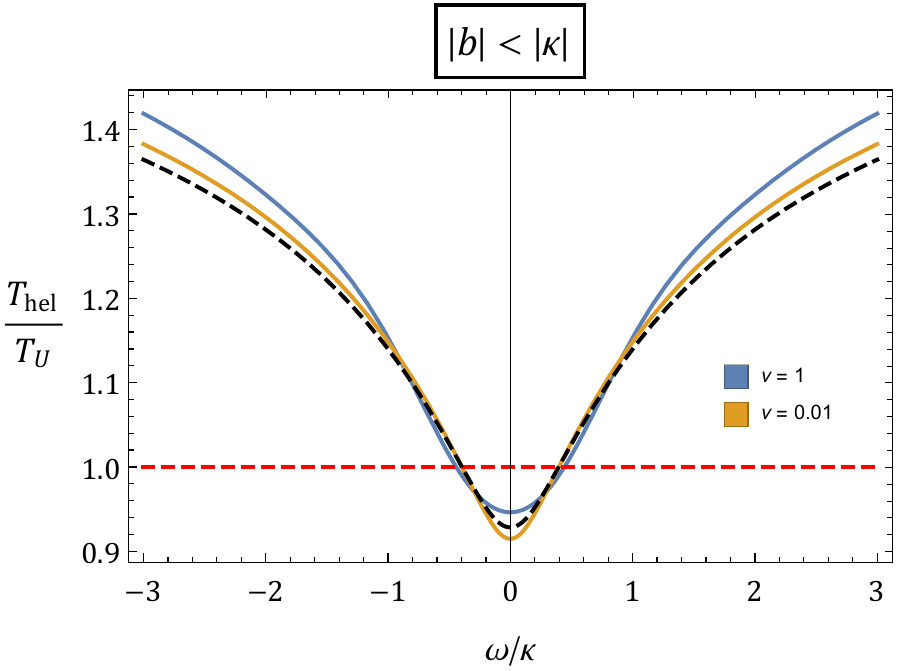}}\hspace{0.5cm}
    \subfigure{\includegraphics[width=0.45\textwidth]{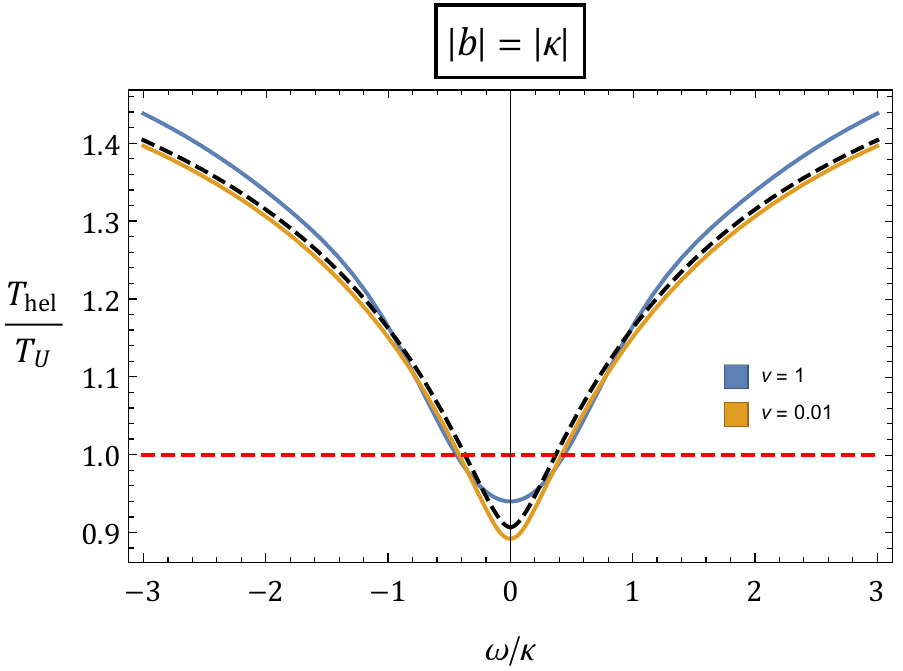}} 
    \subfigure{\includegraphics[width=0.45\textwidth]{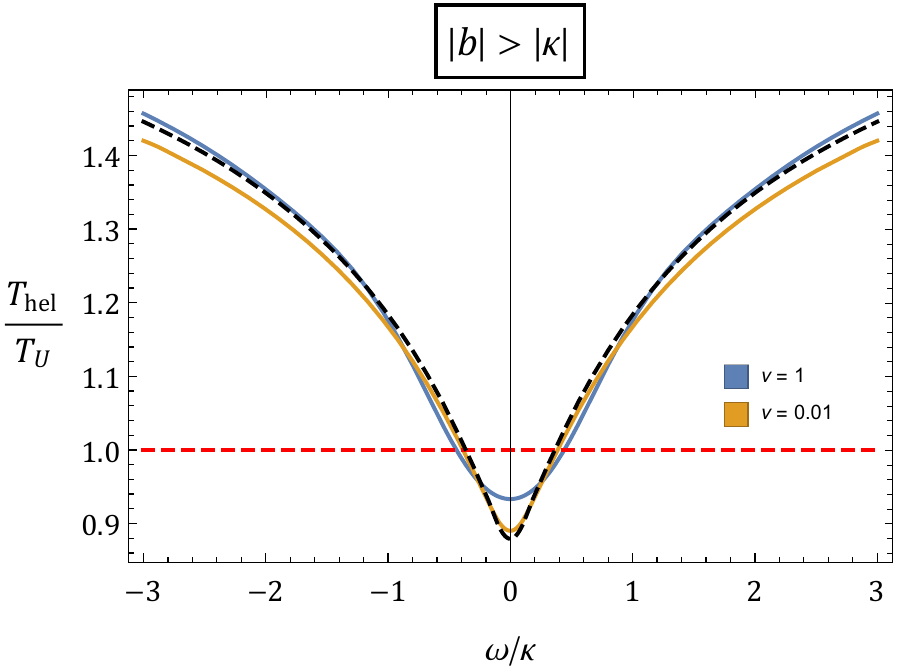}}
    \caption{Effective temperatures along the Hypertor (helix) worldline. The Unruh temperature, $T_{\rm U}$, is indicated by a red dotted line. For $|b| < |\kappa|$, the Infrator (catenary) temperature, $T_{\rm cat}$, is indicated by a black dotted line. For $|b| = |\kappa|$, the Parator (cusped) motion temperature, $T_{\rm cus}$, is indicated by a black dotted line. For $|b| > |\kappa|$, the Ultrator (circular) motion temperature, $T_{\rm cus}$, is indicated by a black dotted line. In each case, at small hypertorsion the Hypertor temperature curve asymptote the black dotted lines.
%    The asymptotic limits along the Hypertor motion are marked with a dotted line, while the other two lines in each graph indicates the Hypertor motion with a large ($\nu=1$) and small ($\nu=0.01$) hypertorsion value.
}
    \label{fig:Eq26}
\end{figure}

In Appendix \ref{App:Helix} we show that the following asymptotic limits for the effective temperatures along the Hypertor motion hold
\begin{subequations}
\label{HelLimits}
\begin{align}
\lim_{\nu \to 0} T_{\rm hel}(\omega) & =  T_{\rm cat}(\omega)  ,   &|b| & < | \kappa |, \\
\lim_{\nu \to 0} T_{\rm hel}(\omega) & = T_{\rm cus}(\omega),   &|b| & = | \kappa |\\
\lim_{\nu \to 0} T_{\rm hel}(\omega) & =T_{\rm cir}(\omega),  &|b|& > | \kappa |.
\end{align}
\end{subequations}

Note that as $\nu \to 0$, the minimum speed for the helix motion tends to $v_{\rm min} = \frac{\kappa}{b}$ in the circular-motion limit with $\gamma = \alpha$, $r = \frac{\beta}{R_-}$, and $\Omega = \frac{R_-}{\alpha}$. The $\nu \to 0$ limit can be compared directly with the circular trajectory, Eq.~\eqref{circtraj}, in the frame connected by the translation $x' = x - \frac{\alpha}{R_+}$ . 

In the catenary-motion limit as $\nu \to 0$, we have that the minimum speed tends to $v_{\min} = \frac{b}{\kappa}$ with $\sigma^2 = R_+^2$. The $\nu \to 0$ limit can be compared directly with the catenary trajectory, Eq.~\eqref{circtraj}, in the frame connected by the translation $y' = y - \frac{\beta}{R_-}$ . 

In the cusped-motion limit, $v_{\rm min}$ is a high-speed limit in the rest frame.  We can see that in the case $b^2 = k^2$, performing the boost $t' = \gamma(t - v_{\rm min} z)$ and $z' = \gamma \left(z - v_{\rm min} t \right)$ along the (t,z) plane with $v_{\rm min} = \frac{\beta}{\alpha}$, and the rotation with translations $x' = \frac{1}{1+v_{\rm min}^2}\left(x - v_{\rm min} y \right) - \frac{5}{4 \kappa}$ and $y' = \frac{1}{1+v_{\rm min}^2}\left(y + v_{\rm min} x \right) - \frac{1}{\nu}$ along the $(x, y)$-plane, one has that
\begin{subequations}
\begin{align}
t'(\tau) & = \gamma \left(\frac{\alpha}{R_+}\sinh{R_+ \tau} - \frac{\beta}{\alpha} \frac{\beta}{R_-}\sin{R_- \tau}   \right) \xrightarrow[\nu \to 0]{} \tau + \frac{ 1}{6 } \kappa ^2 \tau ^3, \\
x'(\tau) & = \frac{1}{1 + \frac{\beta^2}{\alpha^2}} \left( \frac{\alpha}{R_+}\cosh{R_+ \tau} - \frac{\beta}{\alpha} \frac{\beta}{R_-}\cos{R_- \tau}  \right) - \frac{5}{4 \kappa} \xrightarrow[\nu \to 0]{}   \frac{1}{2} \kappa \tau^2, \\
y'(\tau) & =  \frac{1}{1 + \frac{\beta^2}{\alpha^2}} \left( \frac{\beta}{R_-}\cos{R_- \tau} + \frac{\beta}{\alpha} \frac{\alpha}{R_+}\cosh{R_+ \tau}  \right) - \frac{1}{\nu}\xrightarrow[b \to 0]{} 0 , \\
z'(\tau) & = \gamma \left(\frac{\beta}{R_-}\sin{R_- \tau} - \frac{\beta}{\alpha}  \frac{\alpha}{R_+}\sinh{R_+ \tau}  \right) \xrightarrow[\nu \to 0]{} -\frac{ 1}{6 } \kappa^2 \tau^3.
\end{align}
\end{subequations}
which (upon performing a reflection along the $z$-axis) gives the cusped trajectory Eq.~\eqref{cusp} in the new frame.

%the helix motion Eq.~\eqref{hyptraj} tends to circular motion Eq.~\eqref{circtraj}, catenary motion Eq.~\eqref{cattraj} or cusp motion Eq.~\eqref{cusp}, but rather that the temperatures registered along the helix motion at low hypertorsion tend to the temperature registered along the circular, catenary or cusp motions, cf. Eq.~\eqref{HelLimits}.

%In view of Eq. \eqref{Hypertorvmin} (and the comment in the paragraph below that equation), the limits presented in Eq. \eqref{HelLimits} represent rest-frame high-speed limits along the Hypertor motion. 
The limits Eq.~\eqref{HelLimits} are verified numerically in Fig. \ref{fig:Eq26}, where we also explore the large hypertosion regime ($\nu = 1$).

\begin{figure}[t!]
\begin{center}
{\rotatebox{0}{\includegraphics[width=5.0in]{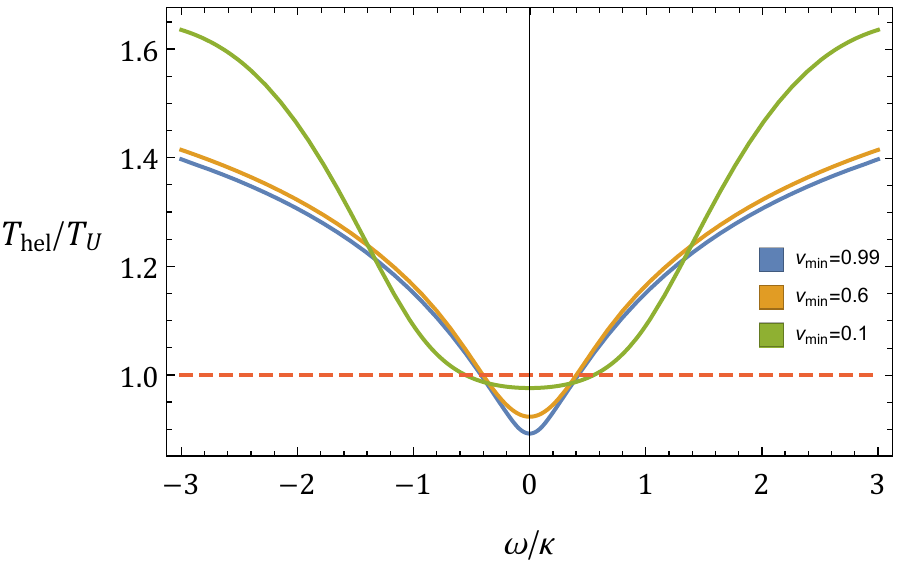}}} 
\caption{\label{fig:WL_Hypertor2} 
Effective temperature along the Hypertor (helix) worldline for $\kappa = b = 1$ and the speed varies as a function of hypertorsion obeying Eq.~ \eqref{Hypertorvmin}. Near the speed of light, for $v_{\rm min} = 0.99$, $\nu \approx 0.010$. For $v_{\rm min} = 0.6$, $\nu \approx 0.516$ and for $v_{\rm min} = 0.6$, $\nu \approx 2.846$ . The Unruh temperature, $T_{\rm U}$, is indicated by a red dashed line.}
%
%When minimum velocity of motion is close to  speed of light, $v_{\rm min} = 0.99$ , hypertorsion is equal to 0.01005. In the second and third case the hypertorsion is  equal to 0.51639 and 2.84604 respectively.    
 
\end{center}
\end{figure}

\begin{figure}[t!]
\begin{center}
\includegraphics[scale=0.55]{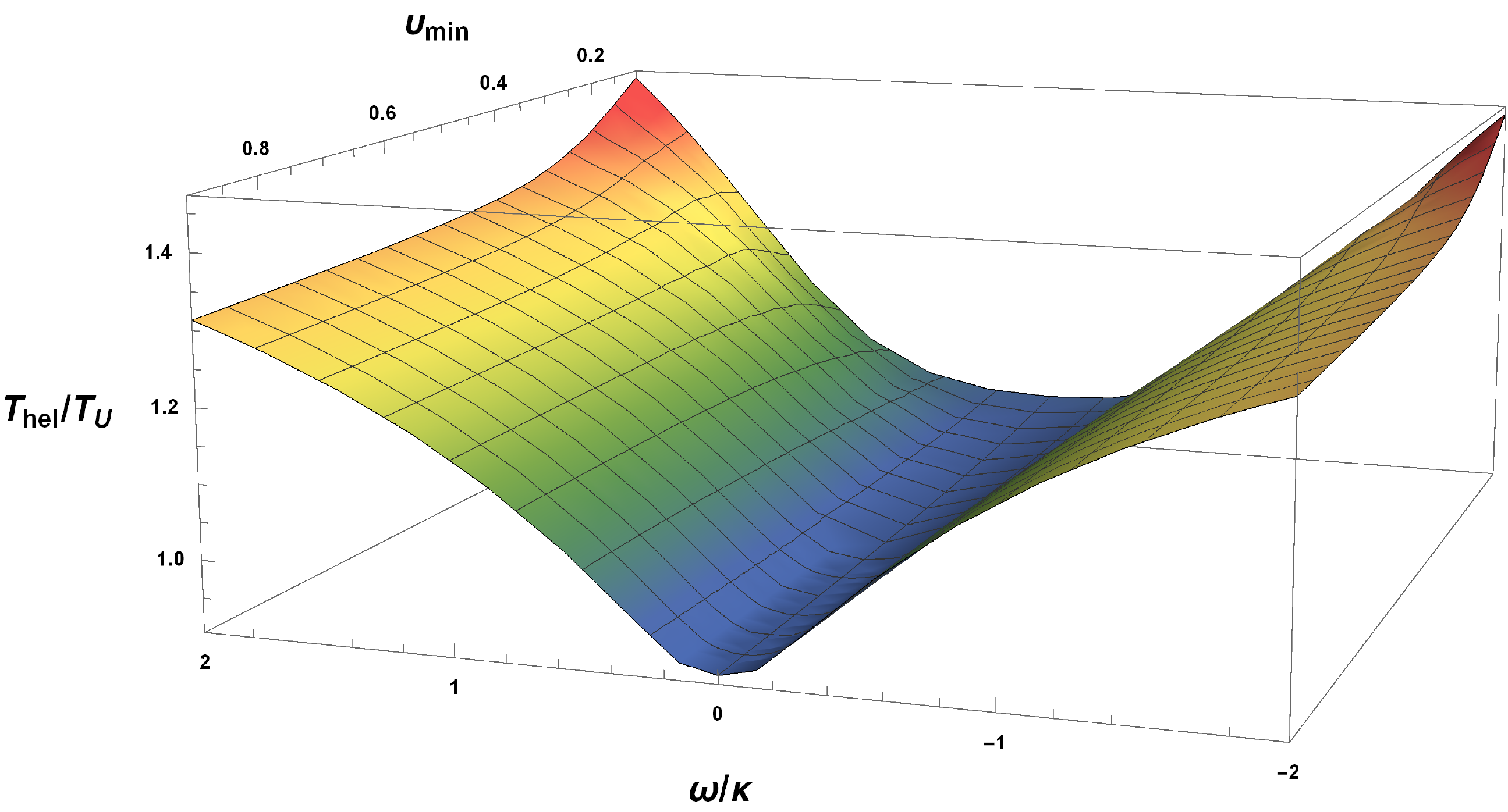}
\caption{\label{fig:3Dhelix} 
The effective temperature scaled to the Unruh temperature as a function of $\omega/\kappa$ and the minimum velocity $0.1\leq\upsilon_{min}\leq0.9$ in the case of hypertor motion with  $\kappa = b = 1$. 
The coldest temperature  $T_{\rm hel}/T_U \approx 0.90$ is  registered  by a detector moving with velocity $\upsilon=0.9$, in the limit $\omega/\kappa\to0$.}
\end{center}
\end{figure}

Further numerics are presented in Fig.~\ref{fig:WL_Hypertor2} and Fig.~\ref{fig:3Dhelix}, where we show that at different values of $v_{\rm min}$ the effective temperatures has a qualitatively similar behaviour. For concreteness, we have chosen to study numerically cases in which $\kappa = b$.

%\section{Effective temperatures at low and high frequencies}
%\label{Sec:Freq}

%The temperatures recorded along the stationary worldlines described in Table \ref{table:traj} will depend generally on the transition frequency difference of the Unruh-DeWitt detector, the notable exception being the constant, Unruh temperature. In this section, we study the low and high-frequency behaviour of these temperatures.

\newpage

\section{Conclusions}\label{conclusions}

We have studied by a combination of numerical and analytic techniques the temperatures recorded by Unruh-DeWitt detectors along all of the stationary worldlines in Minkowski spacetime with non-trivial uniform acceleration classified in \cite{Letaw:1980yv}. %We have investigated the characteristic Unruh frequencies of the motions, their asymptotic values, and temperature dependent behavior. 

%\benito{What is the characteristic Unruh frequency? I have only seen this expression used in the cusped case.} \mike{True, just the intersection with $T=1$. The number itself has not been enlightening, as it is clearly parameter dependent (for traj. other than cusp) and non-analytic.  So it does not appear so fundamental, as might be expected from the cusp result.} \benito{That's interesting. Were people expecting this to be enlightening or important? If so, we should write a comment on it not being straightforwardly relevant for more general stationary trajectories here in the conclusions.}

From our numerical analysis, we have found that:

\begin{itemize}
\item Torsion affects the slope of the effective temperature with respect to frequency by a subsequent increase, at least in the case of ultrator and infrator.
\item Torsion cools the system at zero frequency for all stationary worldlines relative to the Unruh temperature.   
\item Frequency monotonically increases the temperature of the system.
\end{itemize}

From an analytic point of view, we have studied which limits connect the temperatures along different trajectories, as is summarized in Fig.~\ref{fig:Hierarchy}.  These limits are physically approached at high or low speeds for the detectors. We have discussed that in the cases of high speed, one can interpret the ultra-relativistic limit by performing Poincar\'e transformations that depend on $\kappa$, $b$ and $\nu$ (and diverging in a controlled fashion in the appropriate limit if necessary), which allow one to identify different worldlines in different reference frames.

We feel that these results may have important application in the study of the temperature of radiation in the context of quantum field theory in flat spacetime, both theoretically and experimentally. In particular, from an experimental viewpoint, the circular and ``near-circular" helix motions can be implemented in an experiment confined to a ``small" lab region, making them attractive candidates for the measurement of the (generalized) Unruh effect. We must say that for the helix motion the size of the table-top experiment will be proportional to the time scale of the experiment, with a more favorable proportionality factor (depending on the parameter $\kappa$, $b$ and $\nu$) in the near-circular regime.

We have aimed for this work to be thorough, in the sense that no other uniformly accelerated trajectories exist to investigate via this effective temperature approach in Minkowski spacetime.  It is our hope that the general statements we were able to make concerning quantum field theory in flat spacetime via the use of these stationary trajectories contribute to the interesting ongoing investigations on acceleration radiation. 

As a final word, we have numerically investigated the coldest temperatures recorded by an Unruh-DeWitt detector along the stationary worldlines with non-trivial acceleration classified by Letaw \cite{Letaw:1980yv} (at $\omega = 0$), but from an experimental viewpoint, it is appealing to study the {{\emph{hotter}}} temperatures (for large $|\omega|$). This will be the subject of forthcoming work.

\section*{Acknowledgments}
B.A.J-.A. and D.M. thank Jorma Louko for a stimulating email exchange and for pointing out the work appearing in Ref. \cite{Gooding:2020} to us. B.A.J.-A.  is  supported by the Direcci\'on  Genereral  de  Asuntos  del  Personal  Acad\'emico, Universidad Nacional Autnoma de M\'exico through a DGAPA-UNAM postdoctoral fellowship held at Mexico City, and acknowledges additional support from the Sistema Nacional de Investigadores,  Consejo  Nacional de Ciencia y Tecnolog\'ia (SNI-CONACYT),  Mexico. M.R.R.G. and M.T. acknowledge funding from state-targeted program ``Center of Excellence for Fundamental and Applied Physics" (BR05236454) by the Ministry of Education and Science of the Republic of Kazakhstan, as well as funding by the ORAU FY2018-SGP-1-STMM Research Grant No. 090118FD5350 at Nazarbayev University. D. M.'s research is co-financed by Greece and the  European Union (European Social Fund-ESF) through the Operational Programme ``Human Resources Development, Education and Lifelong Learning" in the context of the project ``Reinforcement of Postdoctoral Researchers - 2nd Cycle" (MIS-5033021), implemented by the State Scholarships Foundation (IKY). D.M. acknowledges the hospitality of University of Waterloo and the Institute for Quantum Computing (IQC), where part of the work was carried out.

\appendix

\section{Asymptotic analysis in Sec. \ref{sec:Cir}}
\label{App:Cir}

We are interested in showing that
\begin{subequations}
\label{app:circlimits}
\begin{align}
0 = \left[\lim_{\kappa \to 0} \frac{1}{\omega} \ln \left( \frac{\widehat{\mathcal{W}}^+_{\rm cir}(-\omega)}{\widehat{\mathcal{W}}^+_{\rm cir}(-\omega)} \right)\right]^{-1},  \label{app:circKappa0}\\
T^{-1}_{\rm cus}(\omega) = \lim_{b \to \kappa} \frac{1}{\omega} \ln \left( \frac{\widehat{\mathcal{W}}^+_{\rm cir}(-\omega)}{\widehat{\mathcal{W}}^+_{\rm cir}(-\omega)} \right). \label{app:circKappab}
\end{align}
\end{subequations}

For stationary detectors, one can write the Fourier transform of the pullback of the Wightman function as \cite{ Juarez-Aubry:2014jba, Louko:2006zv}
\begin{equation}
\widehat{\mathcal{W}}^+(\omega) = - \frac{\omega}{2 \pi} \Theta(-\omega) + 2 \int_0^\infty \dd s \cos (\omega s ) \left( \mathcal{W}^+(s) + \frac{1}{4 \pi^2 s^2} \right)
\label{W-hatStat}
\end{equation}

Eq. \eqref{W-hatStat} holds in particular if $\mathcal{W}^+(s) = \mathcal{W}^+_{\rm cir}(s)$. Notice that pointwise
 \begin{subequations}
\begin{align}
\lim_{\kappa \to 0} \mathcal{W}^+_{\rm cir}(s) = \frac{1}{4 \pi^2 s^2},\\
\lim_{b \to \kappa} \mathcal{W}^+_{\rm cir}(s) = \mathcal{W}^+_{\rm cus}(s).
\end{align}
\end{subequations}

Thus, if we can apply dominated convergence to the integral expression
\begin{align}
I(s) := 2 \int_0^\infty \dd s \cos (\omega s ) \left( \mathcal{W}^+_{\rm cir}(s) + \frac{1}{4 \pi^2 s^2} \right),
\end{align}and take the relevant limits inside the integral, Eq. \eqref{app:circlimits} follow.

We begin by studying the $\kappa \to 0$ limit. The pullback of the Wightman function can be written in terms of the curvature, $\kappa$, and torsion, $b$, as
\begin{align}
\mathcal{W}^+_{\rm cir}(s) = \frac{1}{4 \pi^2}\frac{\left(b^2-\kappa ^2\right)^2}{-b^2 s^2 \left(b^2- \kappa^2\right) +2 \kappa ^2\left(1-  \cos \left(s \sqrt{b^2-\kappa ^2}\right) \right)}.
\label{W+cirbkappa}
\end{align}

We now show that 
\begin{align}
\partial_\kappa \left( \mathcal{W}^+_{\rm cir}(s) + \frac{1}{4 \pi^2 s^2} \right) \leq 0.
\label{MonotoneCirc}
\end{align}

Assume $0 < \kappa \leq |b|$. The case for $-|b| \leq \kappa \leq 0$ is analogous. It is easy to see that the numerator of Eq. \eqref{W+cirbkappa} is a non-increasing function of $\kappa$.  The denominator of Eq. \eqref{W+cirbkappa} satisfies
\begin{align}
& \frac{1}{2 \kappa}\partial_\kappa \left( -b^2 s^2 \left(b^2- \kappa^2\right) +2 \kappa ^2\left(1-  \cos \left(s \sqrt{b^2-\kappa ^2}\right) \right) \right) \nonumber \\
& =  b^2 s^2-\frac{\kappa ^2 s \sin \left(s \sqrt{b^2-\kappa ^2}\right)}{\sqrt{(b-\kappa ) (b+\kappa )}}-2 \cos \left(s \sqrt{b^2-\kappa ^2}\right)+2 \nonumber \\
& \geq b^2 s^2\left( 1 -\frac{\sin \left(s \sqrt{b^2-\kappa ^2}\right)}{s \sqrt{(b-\kappa ) (b+\kappa )}} \right) \geq 0,
\end{align}so it is a non-decreasing function of $\kappa$. Thus, Eq. \eqref{MonotoneCirc} follows. Hence
\begin{align}
\mathcal{W}^+_{\rm cus}(s) + \frac{1}{4 \pi^2 s^2}  \leq \mathcal{W}^+_{\rm cir}(s) + \frac{1}{4 \pi^2 s^2}  \leq 0,
\end{align}
and it follows that
\begin{align}
2 \left| \cos (\omega s ) \left( \mathcal{W}^+_{\rm cir}(s) + \frac{1}{4 \pi^2 s^2} \right) \right| \leq 2 \left| \left( \mathcal{W}^+_{\rm cus}(s) + \frac{1}{4 \pi^2 s^2} \right) \right|,
\label{cirbound}
\end{align}
from which dominated convergence follows. Bound Eq.~\eqref{cirbound} holds also if $\kappa \leq 0$.

For Eq. \eqref{app:circKappab}, the limit $b  \to \kappa$ in Eq. \eqref{app:circKappab} can be exchanged  by the limit $\kappa \to b$ on the right-hand side, and dominated convergence also follows also from bound Eq.~\eqref{cirbound}.

\section{Asymptotic analysis in Sec. \ref{Sec:Infrator}}
\label{App:Cat}

We are interested in showing that
\begin{subequations}
\label{cusplimits}
\begin{align}
\frac{2 \pi}{\kappa} = \lim_{b \to 0} \frac{1}{\omega} \ln \left( \frac{\widehat{\mathcal{W}}^+_{\rm cat}(-\omega)}{\widehat{\mathcal{W}}^+_{\rm cat}(-\omega)} \right), \\
T^{-1}_{\rm cus}(\omega) = \lim_{b \to \kappa} \frac{1}{\omega} \ln \left( \frac{\widehat{\mathcal{W}}^+_{\rm cat}(-\omega)}{\widehat{\mathcal{W}}^+_{\rm cat}(-\omega)} \right).
\end{align}
\end{subequations}

For obtaining the $b \to 0$ and $b \to \kappa$ asymptotics, we will apply the dominated convergence theorem. 

Inserting ${\mathcal{W}}^+ = {\mathcal{W}}^+_{\rm cat}$ in  Eq. \eqref{W-hatStat}, we have that
\begin{equation}
\widehat{\mathcal{W}}_{\rm cat}^+(\omega) = - \frac{\omega}{2 \pi} \Theta(-\omega) + 2 \int_0^\infty \dd s \cos (\omega s ) \left( \mathcal{W}^+_{\rm cat}(s) + \frac{1}{4 \pi^2 s^2} \right)
\label{W-hatCat}
\end{equation}

The dominated convergence argument follows from noticing that in our case of interest (see Eq. \eqref{W+Infrator}),
\begin{equation}
\partial_b \left( \frac{\sigma^4}{ \left(\sigma^2 b^2 s^2 -4 \left(\sigma^2+b^2\right) \sinh (\sigma s/2)^2\right)} + \frac{1}{ s^2} \right) = -\frac{2 \sigma^4 b \left(\sigma^2 s^2 -4 \sinh^2 (\sigma s/2)\right)}{ \left(\sigma^2 b^2 s^2 -4 \left(\sigma^2+b^2\right) \sinh^2 (\sigma s/2)\right)^2}
\label{Monotone1}
\end{equation}
has fixed sign for all $s \geq 0$ and $0 < |b| < |\kappa|  $. Indeed, if $b > 0$, the right-hand side of Eq.~\eqref{Monotone1} is non-negative and, if $b < 0$, it is non-positive. It follows that
\begin{subequations}
\label{CatBounds}
\begin{align}
& \mathcal{W}^+_{\rm lin}(s) \leq \mathcal{W}^+_{\rm cat}(s) \leq \mathcal{W}^+_{\rm cus}(s), \text{ if } b > 0, \\
& \mathcal{W}^+_{\rm lin}(s) \geq \mathcal{W}^+_{\rm cat}(s) \geq \mathcal{W}^+_{\rm cus}(s), \text{ if } b < 0,
\end{align}
\end{subequations}
Bounds Eq.~\eqref{CatBounds} imply that
\begin{align}
 \left| \cos (\omega s ) \left( \mathcal{W}^+_{\rm cat}(s) + \frac{1}{4 \pi^2 s^2} \right) \right| & \leq \left| \left( \mathcal{W}^+_{\rm cat}(s) + \frac{1}{4 \pi^2 s^2} \right) \right| \nonumber \\
& \leq  \left| \left( \mathcal{W}^+_{\rm cus}(s) + \frac{1}{4 \pi^2 s^2} \right) \right|  +  \left| \left( \mathcal{W}^+_{\rm lin}(s) + \frac{1}{4 \pi^2 s^2} \right) \right|,
\end{align}where the right-hand side is a $b$-independent, integrable function that provides the dominated convergence argument, from where it follows that the limits of interest ($b \to 0$ and $b \to \kappa$) can be taken inside the integral. Namely,
\begin{subequations}
\begin{align}
\lim_{b \to 0} \widehat{\mathcal{W}}_{\rm cat}(\omega) & = - \frac{\omega}{2 \pi} \Theta(-\omega) + 2 \int_0^\infty \dd s \lim_{b \to 0} \cos (\omega s ) \left( \mathcal{W}^+_{\rm cat}(s) + \frac{1}{4 \pi^2 s^2} \right) \nonumber \\
& =  - \frac{\omega}{2 \pi} \Theta(-\omega) + 2 \int_0^\infty \dd s  \cos (\omega s ) \left( \mathcal{W}^+_{\rm lin}(s) + \frac{1}{4 \pi^2 s^2} \right), \\
\lim_{b \to \kappa} \widehat{\mathcal{W}}_{\rm cat}(\omega) & = - \frac{\omega}{2 \pi} \Theta(-\omega) + 2 \int_0^\infty \dd s \lim_{b \to \kappa} \cos (\omega s ) \left( \mathcal{W}^+_{\rm cat}(s) + \frac{1}{4 \pi^2 s^2} \right) \nonumber \\
& =  - \frac{\omega}{2 \pi} \Theta(-\omega) + 2 \int_0^\infty \dd s  \cos (\omega s ) \left( \mathcal{W}^+_{\rm cus}(s) + \frac{1}{4 \pi^2 s^2} \right),
\end{align}
\end{subequations}
which yield Eq. \eqref{cusplimits}.

\section{Asymptotic analysis of Sec. \ref{Sec:Hypertor}}
\label{App:Helix}

We are interested in showing that
\begin{subequations}
\label{hel-limits}
\begin{align}
T^{-1}_{\rm cat}(\omega) & = \lim_{\nu \to 0} \frac{1}{\omega} \ln \left( \frac{\widehat{\mathcal{W}}^+_{\rm hel} (-\omega)}{\widehat{\mathcal{W}}^+_{\rm hel}(-\omega)} \right),  & |b|& < | \kappa |, \\
T^{-1}_{\rm cus}(\omega)&  = \lim_{\nu \to 0} \frac{1}{\omega} \ln \left( \frac{\widehat{\mathcal{W}}^+_{\rm hel}(-\omega)}{\widehat{\mathcal{W}}^+_{\rm hel}(-\omega)} \right),  & |b| &= | \kappa |\\
T^{-1}_{\rm cir}(\omega)&  = \lim_{\nu \to 0} \frac{1}{\omega} \ln \left( \frac{\widehat{\mathcal{W}}^+_{\rm hel}(-\omega)}{\widehat{\mathcal{W}}^+_{\rm hel}(-\omega)} \right), & |b| &> | \kappa |.
\end{align}
\end{subequations}

First, note that it holds pointwise that
\begin{subequations}
\label{Helpointwise}
\begin{align}
\lim_{\nu \to 0} \mathcal{W}^+_{\rm hel} (s) & = \mathcal{W}^+_{\rm cat} (s), & |b|& < | \kappa |, \label{HelCatLim} \\
\lim_{\nu \to 0} \mathcal{W}^+_{\rm hel} (s) & = \mathcal{W}^+_{\rm cus} (s), & |b|& = | \kappa |, \label{HelCuspLim}\\
\lim_{\nu \to 0} \mathcal{W}^+_{\rm hel} (s) & = \mathcal{W}^+_{\rm cir} (s), & |b|& > | \kappa |. \label{HelCirLim}
\end{align}
\end{subequations}

Our strategy is to use dominated convergence (as was done analogously in Appendix \ref{App:Cir} and Appendix \ref{App:Cat}), whereby we can take the $\nu \to 0$ limit inside the integrals on the right-hand side of Eq.  \eqref{hel-limits} and use Eq.  \eqref{Helpointwise} to obtain the results that we seek.

It is easy to see that Def. \eqref{Helix-Parameters} imply that
\begin{subequations}
\label{Helix-Parameters2}
\begin{align}
R^4 & = (\kappa^2 + T^2 + \nu^2)^2 - 4 \kappa^2 T^2, \\
\Delta^2 & = \frac{1}{2} (R^2 + \kappa^2 + T^2 + \nu^2), \\
R_+^2 & = \frac{1}{2} (R^2 + \kappa^2  - T^2 - \nu^2), \\
R_-^2 & = \frac{1}{2} (R^2 - \kappa^2  + T^2 + \nu^2),
\end{align}
\end{subequations}
from which the identities $R^2 \Delta^2 = \Delta^4 - T^2 \kappa^2$ and $R_+^2 R_-^2 = 4 \kappa^2 \nu^2$ can be deduced. Using Eq. \eqref{W-hatStat} we can write using such identities
\begin{align}
\widehat{\mathcal{W}}^+_{\rm hel} (\omega) & = - \frac{\omega}{2 \pi} \Theta(-\omega)  \nonumber \\
& +\frac{1}{2\pi^2} \int_0^\infty \dd s \cos(\omega s) \left( \frac{1}{ s^2}  -\frac{1}{ 4\pi} \frac{4 (\Delta^4 - T^2 \kappa^2) \kappa^2 \nu^2}{ \Delta ^4 R_-^2 \sinh ^2\left(R_+ s/2\right)- \kappa ^2 R_+^2 T^2 \sin ^2\left(R_- s/2\right)}  \right).
\label{W-hatHel}
\end{align}
We notice that the integrand on the right-hand side of Eq. \eqref{W-hatHel} is even in $\nu$. Therefore, the value at $\nu = 0$ must be an extremal point. This can also be verified directly by using the following identities that follow from \eqref{Helix-Parameters2}
\begin{subequations}
\label{Identities}
\begin{align}
\partial_\nu (R^2) & = 2\left( \frac{2 \Delta^2}{R^2}-1 \right) \nu, \\
\partial_\nu (\Delta^2) & = \frac{2 \Delta^2 \nu}{R^2}, & 
\partial_\nu (\Delta^4) & = \frac{4 \Delta^4 \nu}{R^2}, \\
\partial_\nu (R^2_+) & = 2\left( \frac{ \Delta^2}{R^2}-1 \right)\nu, & \partial_\nu (R_+) & = \frac{1}{R_+}\left( \frac{\Delta^2}{R^2}-1 \right) \nu,  \\
\partial_\nu (R^2_-) & = \frac{2 \Delta^2 \nu }{R^2},  & \partial_\nu (R_-) & = \frac{\Delta^2 \nu}{R_- R^2}.
\end{align}
\end{subequations}

It suffices to show that for the term in the integrand inside the brackets in the right-hand side of Eq.~\eqref{W-hatHel} the extremal point is a maximum or a minimum uniformly in $s$ for $\nu \in [-\nu, \nu_0]$, where $\nu_0$ is some fixed positive number to be able to apply dominated convergence by a strategy similar to the one applied in Appendix \ref{App:Cat}.

Let us first study the case $|b| < |\kappa|$.  Let us write
\begin{equation}
I(\nu, s) :=\frac{1}{ s^2}  -\frac{1}{ 4\pi} \frac{4 (\Delta^4 - T^2 \kappa^2) \kappa^2 \nu^2}{ \Delta ^4 R_-^2 \sinh ^2\left(R_+ s/2\right)- \kappa ^2 R_+^2 T^2 \sin ^2\left(R_- s/2\right)}.
\end{equation}
It can be shown that
\begin{align}
\partial_\nu^2 I(\nu, s) & = \left\{\kappa ^2 \left(\kappa ^2-T^2\right) \left(\kappa ^2+\sqrt{\left(T^2-\kappa ^2\right)^2}+T^2\right) \right\} \nonumber \\
& \times \left\{2 \pi  T^2 \left(\kappa ^2 \left(\kappa ^2+\sqrt{\left(\kappa ^2-T^2\right)^2}-T^2\right) \sin ^2\left(\frac{s \sqrt{-\kappa ^2+\sqrt{\left(T^2-\kappa ^2\right)^2}+T^2}}{2 \sqrt{2}}\right) \right. \right. \nonumber \\
&  \left. \left. +T^2 \left(-\kappa ^2+\sqrt{\left(\kappa ^2-T^2\right)^2}+T^2\right) \sinh ^2\left(\frac{s \sqrt{\kappa ^2+\sqrt{\left(T^2-\kappa ^2\right)^2}-T^2}}{2 \sqrt{2}}\right)\right) \right\}^{-1} \label{ddnuI}
\end{align}
for $|b| < |\kappa|$, the limit $\partial_\nu^2 I(\nu, s)$ diverges with the sign of the function
\begin{align}
h(s) := \kappa^2\left(\kappa^2 - T^2\right) s^2 - 2 T^2+2 T^2 \cosh \left(s \sqrt{\kappa ^2-T^2}\right).
\end{align}
It can be shown by elementary methods that $h(s)\geq 0$ for $s \geq 0$, %. Indeed, $h(0) = 0$ and
%\begin{align}
%h'(s) & = 2 \kappa^2(\kappa^2 - T^2) s+2 T^2 \sqrt{\kappa ^2-T^2} \sinh \left(s \sqrt{\kappa ^2-T^2}\right) \geq 0
%\end{align}
%for $s \geq 0$, 
which implies that $I(0,s)$ is a local minimum uniformly in $s$. Therefore, for $-\nu_0 \leq \nu \leq \nu_0$, the following bound holds for all $s \geq 0$ in the case $|b| < |\kappa|$
\begin{align}
&  \left| \cos(\omega s) \left(\mathcal{W}^+_{\rm hel}(s)  +\frac{1}{4 \pi^2 s^2}\right) \right| \leq \left| \left(\mathcal{W}^+_{\rm hel}(s)  +\frac{1}{4 \pi^2 s^2}\right) \right|\leq \left| \left(\mathcal{W}^+_{\rm cat}(s)  +\frac{1}{4 \pi^2 s^2}\right) \right| \nonumber \\
& + \frac{1}{4 \pi^2} \left| \cos(\omega s) \left( \frac{1}{ s^2}  -\frac{1}{ 4\pi} \frac{4 ((\Delta_0)^4 - T^2 \kappa^2) \kappa^2 (\nu_0)^2}{ (\Delta_0) ^4 (R_{-0})^2 \sinh ^2\left((R_{+0}) s/2\right)- \kappa ^2 (R_{+0})^2 T^2 \sin ^2\left((R_{-0}) s/2\right)}  \right) \right|,
\end{align}
where $\Delta_0$ and $R_{\pm 0}$ are the values of $\Delta$ and $R_\pm$ evaluated at $\nu = \nu_0$, which allows us to take the limit inside in
\begin{align}
\lim_{\nu \to 0} \widehat{\mathcal{W}}^+_{\rm hel} (\omega) & = - \frac{\omega}{2 \pi} \Theta(-\omega)  + 2\int_0^\infty \dd s \lim_{\nu \to 0} \cos(\omega s) \left( \mathcal{W}^+_{\rm hel} +\frac{1}{ 4 \pi^2 s^2}  \right) \nonumber \\
& = - \frac{\omega}{2 \pi} \Theta(-\omega) + 2\int_0^\infty \dd s \cos(\omega s) \left( \mathcal{W}^+_{\rm cat} +\frac{1}{ 4 \pi^2 s^2}  \right), \text{ for } |b| < |\kappa|,
\end{align}
from where the limit Eq.~\eqref{HelCatLim} follows.

Assume now that $|b| > |\kappa|$. In this case, $\partial_\nu^2 I(\nu,s)$ is again given by Eq. \eqref{ddnuI}. In the limit $\nu \to 0$ it diverges with the sign of the function
\begin{equation}
g(s):= -2 \kappa ^2-s^2 T^2 \left(T^2-\kappa ^2 \right)+2 \kappa ^2 \cos \left(s \sqrt{T^2-\kappa ^2}\right),
\end{equation}
which can be shown to be non-positive for $s \geq 0$ by elementary methods.
%Once we show that $g(s)$ has fixed sign for all  $s \geq 0$,  dominated converge follows by arguments analogous to the ones in case $|b| < |\kappa|$. This can be achieved by elementary methods. Indeed, $g(0) = 0$, so it suffices to prove that $g'(s)$ has fixed signed for all $s \geq 0$. We have that
%\begin{align}
%g'(s)& = -2s T^2 \left(T^2-\kappa ^2 \right)-2 \kappa ^2 \sqrt{T^2-\kappa ^2} \sin \left(s \sqrt{T^2-\kappa ^2}\right) \nonumber \\
%& = -2s \left(T^2-\kappa ^2 \right)^2 -2 \kappa^2 \sqrt{T^2 - \kappa^2} \left( s \sqrt{T^2-\kappa ^2} + \sin \left(s \sqrt{T^2-\kappa ^2}\right) \right),
%\end{align}
%which is non-positive since $x + \sin x \geq 0$ for $x \geq 0$, 
This implies that $I(0, s)$ is a local maximum for all $s \geq 0$. Dominated convergence follows from this observation, since the bound
\begin{align}
&  \left| \cos(\omega s) \left(\mathcal{W}^+_{\rm hel}(s)  +\frac{1}{4 \pi^2 s^2}\right) \right| \leq \left| \left(\mathcal{W}^+_{\rm hel}(s)  +\frac{1}{4 \pi^2 s^2}\right) \right|\leq \left| \left(\mathcal{W}^+_{\rm cir}(s)  +\frac{1}{4 \pi^2 s^2}\right) \right| \nonumber \\
& + \frac{1}{4 \pi^2} \left| \cos(\omega s) \left( \frac{1}{ s^2}  -\frac{1}{ 4\pi} \frac{4 ((\Delta_0)^4 - T^2 \kappa^2) \kappa^2 (\nu_0)^2}{ (\Delta_0) ^4 (R_{-0})^2 \sinh ^2\left((R_{+0}) s/2\right)- \kappa ^2 (R_{+0})^2 T^2 \sin ^2\left((R_{-0}) s/2\right)}  \right) \right|,
\end{align}holds for all $s \geq 0$ and $|b| > |\kappa|$. Thus,
\begin{align}
\lim_{\nu \to 0} \widehat{\mathcal{W}}^+_{\rm hel} (\omega) & = - \frac{\omega}{2 \pi} \Theta(-\omega)  + 2\int_0^\infty \dd s \lim_{\nu \to 0} \cos(\omega s) \left( \mathcal{W}^+_{\rm hel} +\frac{1}{ 4 \pi^2 s^2}  \right) \nonumber \\
& = - \frac{\omega}{2 \pi} \Theta(-\omega) + 2\int_0^\infty \dd s \cos(\omega s) \left( \mathcal{W}^+_{\rm cir} +\frac{1}{ 4 \pi^2 s^2}  \right),  \text{ for } |b| < |\kappa|
\end{align}
from where the limit Eq.~\eqref{HelCirLim} follows.

Assume now that $|b| = |\kappa|$. In this case

\begin{align}
\partial^2_\nu I(0,s) = -\frac{2 (\kappa  s \sin (\kappa  s)+4 \cos (\kappa  s)-4)}{\pi  (\cos (\kappa  s)+4 \cosh (\kappa  s)-5)^2}.
\label{SecondDerivativeI}
\end{align}
%We proceed to analyze the sign of
%\begin{align}
%n(r) := r \sin (r)+4 \cos (r)-4
%\end{align}
%for $r \geq 0$ if $\kappa > 0$ and $r \leq 0$ if $\kappa < 0$. This can be done by elementary methods. First, note that $n(0) = 0$. Second, we have that
%\begin{align}
%n'(r) = r \cos (r)-3 \sin (r),
%\end{align}
%only contains real roots at $r = 0$. This means that $r = 0$ is the only extremal point of $n$. Analyzing the second derivative we see that $r = 0$ is a maximum, from where it follows that $n(r) \leq 0$ for all $r \in \mathbb{R}$, 

We can deduce that $I(0,s)$ is a minimum for all $s \geq 0$ from eq. \eqref{SecondDerivativeI}. Dominated convergence follows by arguments similar to the ones discussed in the cases $|b| > |\kappa|$ and $|b| > |\kappa|$, whereby 
\begin{align}
\lim_{\nu \to 0} \widehat{\mathcal{W}}^+_{\rm hel} (\omega) & = - \frac{\omega}{2 \pi} \Theta(-\omega)  + 2\int_0^\infty \dd s \lim_{\nu \to 0} \cos(\omega s) \left( \mathcal{W}^+_{\rm hel} +\frac{1}{ 4 \pi^2 s^2}  \right) \nonumber \\
& = - \frac{\omega}{2 \pi} \Theta(-\omega) + 2\int_0^\infty \dd s \cos(\omega s) \left( \mathcal{W}^+_{\rm cus} +\frac{1}{ 4 \pi^2 s^2}  \right),  \text{ for } |b| = |\kappa|
\end{align}
from where the limit Eq.~\eqref{HelCuspLim} follows.

\end{document}